\documentclass[twocolumn]{pasj01}

\usepackage{url}
\newenvironment{trueauthors}{\section*{Author contributions}\fontsize{8}{11}\selectfont}{\par}

\begin{document} 
\Received{2017 June 30}
\Accepted{2017 December 6}

\title{Hitomi Observations of the LMC SNR N132D: Highly
Redshifted X-ray Emission from Iron Ejecta
\thanks{The corresponding authors are Eric D.~Miller, Hiroya Yamaguchi,
Kumiko Nobukawa, Makoto Sawada, Masayoshi Nobukawa, Satoru Katsuda, and
Hideyuki Mori.}}

\author{Hitomi Collaboration,
Felix \textsc{Aharonian}\altaffilmark{1,2,3},
Hiroki \textsc{Akamatsu}\altaffilmark{4},
Fumie \textsc{Akimoto}\altaffilmark{5},
Steven W. \textsc{Allen}\altaffilmark{6,7,8},
Lorella \textsc{Angelini}\altaffilmark{9},
Marc \textsc{Audard}\altaffilmark{10},
Hisamitsu \textsc{Awaki}\altaffilmark{11},
Magnus \textsc{Axelsson}\altaffilmark{12},
Aya \textsc{Bamba}\altaffilmark{13,14},
Marshall W. \textsc{Bautz}\altaffilmark{15}
% ...
%
%
}

\altaffiltext{1}{Dublin Institute for Advanced Studies, 31 Fitzwilliam Place, Dublin 2, Ireland}
\altaffiltext{2}{Max-Planck-Institut f{\"u}r Kernphysik, P.O. Box 103980, 69029 Heidelberg, Germany}
\altaffiltext{3}{Gran Sasso Science Institute, viale Francesco Crispi, 7 67100 L'Aquila (AQ), Italy}
\altaffiltext{4}{SRON Netherlands Institute for Space Research, Sorbonnelaan 2, 3584 CA Utrecht, The Netherlands}
\altaffiltext{5}{Institute for Space-Earth Environmental Research, Nagoya University, Furo-cho, Chikusa-ku, Nagoya, Aichi 464-8601}
\altaffiltext{6}{Kavli Institute for Particle Astrophysics and Cosmology, Stanford University, 452 Lomita Mall, Stanford, CA 94305, USA}
\altaffiltext{7}{Department of Physics, Stanford University, 382 Via Pueblo Mall, Stanford, CA 94305, USA}
\altaffiltext{8}{SLAC National Accelerator Laboratory, 2575 Sand Hill Road, Menlo Park, CA 94025, USA}
\altaffiltext{9}{NASA, Goddard Space Flight Center, 8800 Greenbelt Road, Greenbelt, MD 20771, USA}
\altaffiltext{10}{Department of Astronomy, University of Geneva, ch. d'\'Ecogia 16, CH-1290 Versoix, Switzerland}
\altaffiltext{11}{Department of Physics, Ehime University, Bunkyo-cho, Matsuyama, Ehime 790-8577}
\altaffiltext{12}{Department of Physics and Oskar Klein Center, Stockholm University, 106 91 Stockholm, Sweden}
\altaffiltext{13}{Department of Physics, The University of Tokyo, 7-3-1 Hongo, Bunkyo-ku, Tokyo 113-0033}
\altaffiltext{14}{Research Center for the Early Universe, School of Science, The University of Tokyo, 7-3-1 Hongo, Bunkyo-ku, Tokyo 113-0033}
\altaffiltext{15}{Kavli Institute for Astrophysics and Space Research, Massachusetts Institute of Technology, 77 Massachusetts Avenue, Cambridge, MA 02139, USA}
% ...

\email{milleric@mit.edu}


\maketitle

\begin{abstract}
We present Hitomi observations of N132D, a young, X-ray bright, O-rich
core-collapse supernova remnant in the Large Magellanic  Cloud (LMC).
Despite a very short observation of only 3.7 ks, the Soft X-ray
Spectrometer (SXS) easily detects the line complexes of highly ionized S K
and Fe K with 16--17 counts in each.  The Fe feature is measured for the
first time at high spectral resolution.  Based on the plausible assumption
that the Fe K emission is dominated by He-like ions, we find that the
material responsible for this Fe emission is highly redshifted at
$\sim$\,800 km~s$^{-1}$ compared to the local LMC interstellar medium
(ISM), with a 90\% credible interval of 50--1500
km~s$^{-1}$ if a weakly informative prior is placed on possible line
broadening. This indicates (1) that the Fe emission arises from the
supernova ejecta, and (2) that these ejecta are highly
asymmetric, since no blue-shifted component is found.  The S K velocity is
consistent with the local LMC ISM, and is likely from swept-up ISM
material.  These results are consistent with spatial mapping that shows the
He-like Fe concentrated in the interior of the remnant and the S tracing
the outer shell.  The results also show that even with a very small number
of counts, direct velocity measurements from Doppler-shifted lines detected
in extended objects like supernova remnants are now possible.  Thanks to
the very low SXS background of $\sim$\,1 event per spectral resolution
element per 100 ks, such results are obtainable during short pointed or
slew observations with similar instruments.  This highlights the power of
high-spectral-resolution imaging observations, and demonstrates the new
window that has been opened with Hitomi and will be greatly widened with
future missions such as the X-ray Astronomy Recovery Mission (XARM) and
Athena.  
\end{abstract}

\section{Introduction}
\label{sect:intro}

As the main drivers for matter and energy in the Universe, supernova
remnants (SNRs) are excellent laboratories for studying nucleosynthesis
yields and for probing the supernova (SN) engine and dynamics.
Core-collapse SNRs, in particular, address fundamental questions related to
the debated explosion mechanism and the aftermath of exploding a massive
star. 

The mechanism of core-collapse supernova explosions has been one of the
central mysteries in stellar astrophysics.   While one-dimensional
simulations have failed to explode a star, only very recently, successful
explosions of massive stars have been achieved in three-dimensional
simulations invoking convection or standing accretion shock instabilities
(SASI; see \cite{Janka2016} for a recent review).  The ejecta composition
and dynamics as a function of the progenitor star's mass and environment
have formed another puzzle, with predictions largely relying on the
assumption of spherically symmetric models and with yields that vary
depending on metallicity, mass loss, explosion energy, and other
assumptions (e.g., \cite{Nomoto2006,WoosleyHeger2007}). 

Significant progress has been made to answer these central questions,
thanks to high-resolution imaging and spectroscopic mapping of ejecta (in
space and velocity) in core-collapse SNRs, including the oxygen-rich, very
young and bright Cassiopeia~A SNR in our Galaxy
\citep{HwangLaming2012,Grefenstette2014} and more evolved 
SNRs with ejecta signatures such as the O-rich Galactic SNRs G292.2$+$1.8
\citep{Park2007,Kamitsukasa2014} and Puppis~A
\citep{Hwang2008,Katsuda2013}, and the ejecta-dominated SNR W49B
(\cite{Lopez2013a}, \yearcite{Lopez2013b}).  While such observations have
opened a new window to understanding the physics and aftermath of
core-collapse explosions, several complications remain in interpreting the
observations. First, resolving ejecta from the shocked interstellar medium
(ISM) requires fine spectral resolution of extended objects in the X-ray.
Second, there is a strong dependence of the elemental distribution and
plasma state on both the evolutionary stage of the SNR and on the
surrounding environment shaped by the exploded progenitor star.
Mixed-morphology SNRs, expanding into an inhomogeneous medium and often
interacting with molecular clouds, need the additional treatment of
over-ionized (recombining) plasma, as opposed to under-ionized (ionizing)
plasma in the younger remnants or SNRs expanding into a low-density and
homogeneous medium (e.g., \cite{Ozawa2009,Uchida2015}).  The advent of
high-spectral-resolution imaging detectors such as the Soft X-ray
Spectrometer (SXS) aboard Hitomi has promised to revolutionize our
three-dimensional mapping of ejecta dynamics and composition, while
spectroscopically differentiating between shocked ejecta and the shocked
circumstellar/interstellar environment shaped by the progenitor star
\citep{Takahashi16,Hughes2014}.  

A natural early target for Hitomi was N132D, the X-ray brightest SNR in the
LMC, with an age estimated to be $\sim$\,2500\,yr \citep{VogtDopita2011}.
High-velocity ejecta were first detected and studied in optical wavelengths
in N132D \citep{Danziger76,Sutherland95a,Morse95,Morse96}.  Optical/UV
spectra from the Hubble Space Telescope show strong emission of
C/Ne-burning elements (i.e., C, O, Ne, Mg), but little emission from
O-burning elements (i.e., Si, S), leading to an interpretation of a Type Ib
core-collapse supernova origin for this SNR \citep{Blair00}. 

In the X-ray band, the Einstein Observatory made the first observation of
N132D, revealing its clear shell-like morphology \citep{Mathewson83} which
has been interpreted as arising from the SN blast wave expanding within a
cavity produced by the progenitor star's H~\textsc{ii} region
\citep{Hughes1987}.  Einstein also performed the first high-resolution
spectral observations with the Focal Plane Crystal Spectrometer (FPCS),
clearly seeing strong oxygen and other emission lines and obtaining the
first measurements of line flux ratios and constraints on the temperature
and ionization state \citep{Hwang93}.  The following ASCA observations
revealed that elemental abundances of the entire SNR are consistent with
the mean LMC values. This suggests that the X-ray-emitting plasma is
dominated by the swept-up ISM \citep{Hughes98}.  Beppo-SAX detected Fe K
line emission arising from a hot plasma \citep{Favata97}.  High-resolution
X-ray images from XMM-Newton and Chandra have shown that the Fe K-emitting
material is concentrated in the interior of the SNR, contrasting with the
material emitting at softer energies of O, Ne, Mg, Si, S, and Fe L
\citep{Behar01,Canizares01,Borkowski07,XiaoChen08,Plucinsky2016}.  X-ray
emission from O-rich ejecta knots has also been discovered with Chandra,
showing a spatial correlation with the optical O emission
\citep{Borkowski07}.  The centroid and intensity of the Fe K line emission
measured with Suzaku support the core-collapse origin \citep{Yamaguchi14b}.
Very recently, a combined NuSTAR and Suzaku analysis revealed that the hot,
Fe K-emitting plasma is in a recombining state with a large relaxation
timescale of $\sim$\,10$^{12}$\,cm$^{-3}$\,s, implying that the plasma
underwent rapid cooling in the very beginning of its life \citep{Bamba17}.  

N132D is the brightest among all known SNRs in GeV and TeV bands
\citep{Ackermann16,HESS2015}.  The spectral energy distribution from radio
to gamma-rays including synchrotron X-rays detected with NuSTAR suggests
that the gamma-ray emission originates from hadronic processes
\citep{Bamba17}.  The total proton energy required to explain the spectral
energy distribution was derived to be $\sim$\,10$^{50}$\,erg, showing that
N132D is an efficient particle accelerator.

We here present Hitomi observations of N132D. These commissioning phase
observations were expected to explore the emission line structure of the
remnant with exquisite spectral resolution, unprecedented for an extended
object at the energy of Fe K ($\sim$\,6.7 keV).  Unfortunately, due to poor
satellite attitude control during the majority of the observation (see
section \ref{sect:obs} for details), only a short exposure was obtained
with the Hitomi/SXS microcalorimeter.  Nevertheless, owing to the excellent
spectral resolution and gain accuracy of the SXS, we detect spectral
features of strong emission from S, Ar, and Fe, allowing us to investigate
the bulk velocity of the shocked material in this SNR using the Doppler
shift of these emission lines.  We demonstrate the superior capability of
high-resolution spectrometers particularly for low-statistics data, which
provide positive prospects for future observations of distant or faint
objects with future X-ray microcalorimeter missions, like the X-ray
Astronomy Recovery Mission (XARM), Athena \citep{AthenaWP2013}, and
Lynx\footnote{https://wwwastro.msfc.nasa.gov/lynx}.  We also present the
analysis of Soft X-ray Imager (SXI) data, simultaneously obtained from this
observation but with longer exposure (and hence higher statistics) owing to
its wide field of view (FoV). 

This paper is organized as follows. In section \ref{sect:obs}, we describe
the details of the Hitomi observations.  We present spectral analysis of
the SXS and SXI in sections \ref{sect:sxs_analysis} and
\ref{sect:sxi_analysis}, respectively.  We discuss the results in section
\ref{sect:disc} and summarize in section \ref{sect:conc}.  Throughout the
paper, we assume 50~kpc for the distance to the LMC \citep{Westerlund90},
and $v_{\rm helio,LMC} = 275 \pm 4$~km~s$^{-1}$ as the heliocentric
velocity of the LMC ISM immediately surrounding N132D
\citep{VogtDopita2011}.  Heliocentric velocities noted by $v_{\rm helio}$
have been corrected to the Solar System barycentric standard of rest.  The
errors quoted in the text and table represent the 90\% confidence level,
and the error bars given in the spectra represent 68\%
confidence.

\section{Observations and Data Reduction}
\label{sect:obs}

The Hitomi X-ray Observatory was launched in February 2016 and tragically
lost at the end of March \citep{Takahashi16}. During the month of
operation, the SXS successfully demonstrated its in-orbit performance by
achieving an unprecedented spectral resolution ($\Delta E \approx 5$\,eV)
across a broad energy (2--12\,keV) for extended sources
\citep{Kelley16,Porter16}.  This led to accurate determination of the
turbulent velocity of hot plasma in the Perseus Cluster by measuring the
line width of the Fe\,\textsc{xxv} He$\alpha$ fine structure
\citep{Hitomi16}. 

\begin{figure*}[t]
     \begin{center}
          \FigureFile(100,80mm){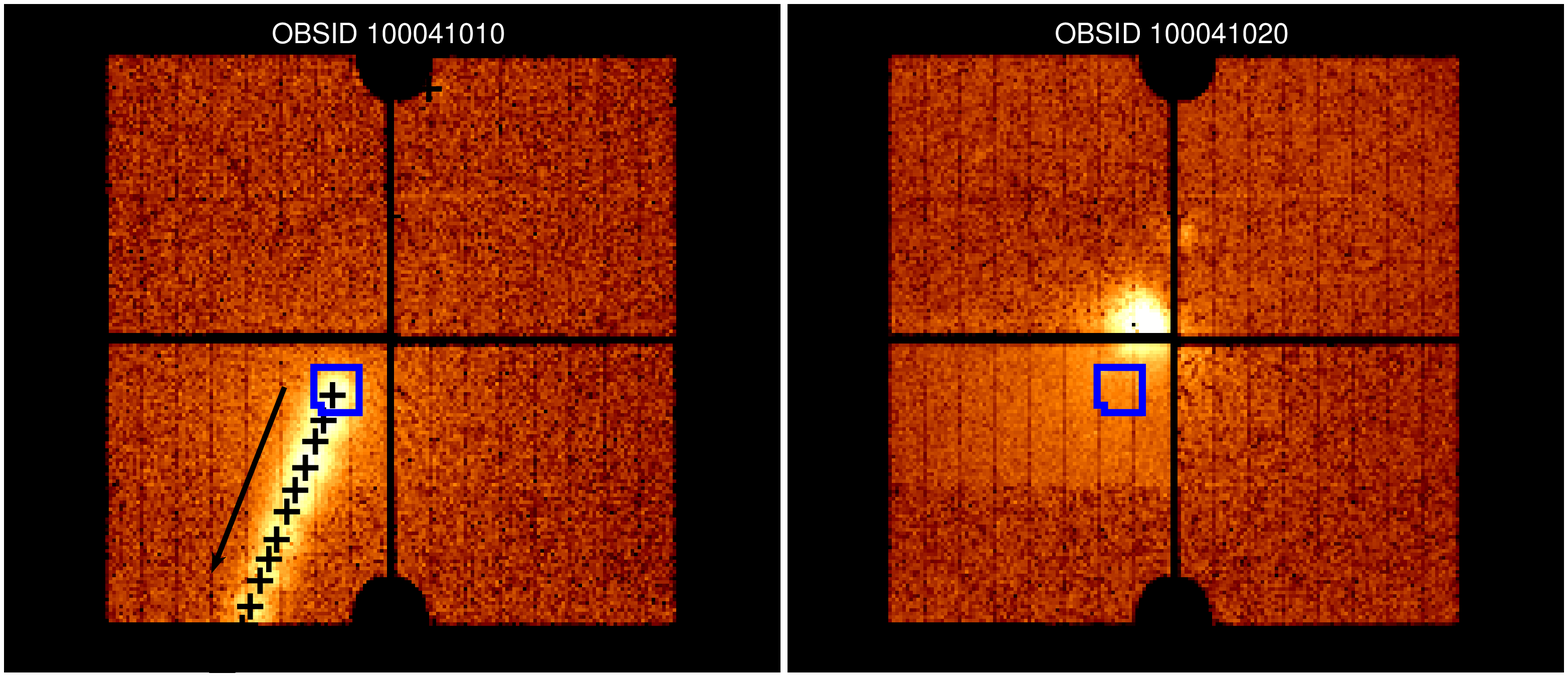}
          \hspace*{5mm}
          \FigureFile(60mm,80mm){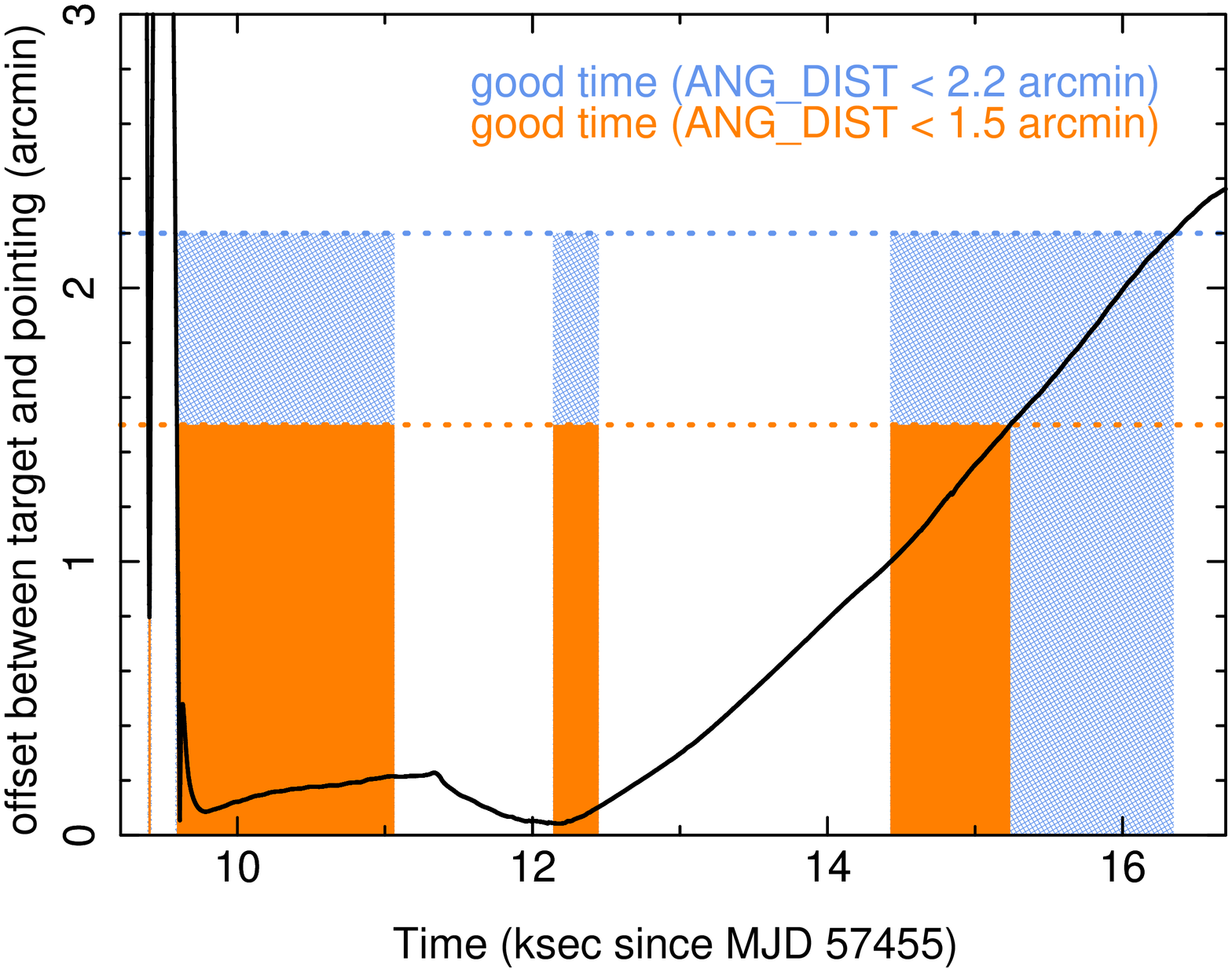}	
     \end{center}
     \caption{
     (Left) SXI images in detector coordinates showing the two OBSIDs used
     in the SXI analysis.  The blue square shows the SXS FoV.  The arrow in
     the left panel shows the direction N132D drifted in the focal plane
     during that OBSID, with black crosses marking the source center in
     intervals of one hour.  N132D was in the SXS FoV for only 3.7 ks of
     OBSID 100041010, and for none of OBSID 100041020.  The remnant was
     still visible in SXI due to that detector's larger FoV.  
     (Right) Attitude of Hitomi during the first 7 ks of OBSID 100041010.
     The solid line shows \texttt{ANG\_DIST}, the angular distance in arcmin
     between the intended pointing and the actual pointing.  Orange bins
     show the good time intervals of the default data filtering, which
     requires \texttt{ANG\_DIST} $<$ 1$.\!\!^\prime$5.  Blue bins show the
     additional $\sim$\,43\% of time added by relaxing this criterion to
     2$.\!\!^\prime$2.  Blank times are excluded because of Earth
     occultation or South Atlantic Anomaly passage.}
     \label{fig:attitude}
     \label{fig:att}
\end{figure*}

After the Perseus observations, Hitomi aimed at the SNR N132D for
performance verification of the SXS and SXI using another line-rich source.
The other detectors, the Hard X-ray Imager (HXI) and Soft Gamma-ray
Detector (SGD), were not yet turned on.  Unfortunately, the satellite
attitude control system lost control about 30 minutes after the observation
started due to problems in the star tracker system, as illustrated in
figure \ref{fig:att}.  Because of this, the SNR drifted out of the
$3\arcmin\times3\arcmin$ SXS FoV and remained out of view for the remainder
of the observation.  Thanks to its larger FoV, the SXI was able to observe
the source during the entire observation. 

As this observation took place during the commissioning phase, several
instrument settings were non-standard compared to expected science
operation.  First, the SXS gate valve was in the closed configuration to
reduce the chance of molecular contamination from spacecraft out-gassing.
The gate valve had a $\sim$\,260 $\mu$m thick Be window to allow
observations while closed, but this absorbed almost all X-rays below
$\sim$\,2 keV and reduced the effective area by $\sim$\,50\% at higher
energies \citep{Eckart2016}.  Thus we limit our SXS analysis to the
2--10~keV regime.  Second, while the SXS was close to thermal equilibrium
at this point in the commissioning phase \citep{Fujimoto2016,Noda2016}, no
on-orbit, full-array energy scale (or gain) calibration had been performed
with the filter-wheel calibration sources.  The Modulated X-ray Source
(MXS; de Vries et al.~2017) was also not available for contemporaneous gain
measurement.  A dedicated calibration pixel that was outside of the
aperture and continuously illuminated by a collimated $^{55}$Fe source
served as the only contemporaneous energy-scale reference, and the
time-dependent scaling required to correct its gain was applied to each
pixel in the array \citep{Porter16b}.   It was well known prior to launch
that the time-dependent gain-correction function for this calibration pixel
generally would not adequately correct the energy scale of the array
pixels.  In particular, the relationship between changes on the calibration
pixel and on the array was not fixed, but rather depended on the
temperatures of various shields and interfaces in the SXS dewar.
Therefore, although the relative drift rates across the array were
characterized during a later calibration with the filter-wheel $^{55}$Fe
source (Eckart et al.~in prep.), changes in SXS cryocooler settings between
the N132D observation and that calibration limit the usefulness of that
characterization.  

In fact, the measured relative gain drift predicts a much larger
energy-scale offset between the final two pointings of the Perseus Cluster
than was actually observed.  Using source-free SXS observations taken
during the period with the same cryocooler settings as the N132D
observation (2016 March 7–-15) in order to circumvent this limitation, we
measured the center of the Mn K$\alpha$ instrumental line (Kilbourne et al.
in prep.), and conclude that the SXS energy scale is shifted by at most
$+1\pm0.5$ eV at 5.9 keV (Eckart et al. in prep.). There are no
sufficiently strong low-energy lines in the same data set, but
extrapolating from Perseus Cluster observations, we estimate a gain shift
of $-2\pm1$ eV at 2 keV (Hitomi Collaboration, in prep. [Perseus cluster
atomic data paper]).  In the filter-wheel $^{55}$Fe data set, errors in the
position of the Mn K$\beta$ line ranged from $-$0.6 to $+$0.2~eV across the
array.  Since this line is at 6.5 keV, less than 1 keV from the Mn
K$\alpha$ reference line, gain errors at other energies further from the
reference may be substantial.  This is especially true in science data, for
which drift of the energy scale can only be corrected via the data from the
calibration pixel.  To be conservative, we use a systematic gain error of
$\pm2$ eV at all energies in the analysis below.

\begin{figure}[t]
     \begin{center}
          \FigureFile(8cm,8cm){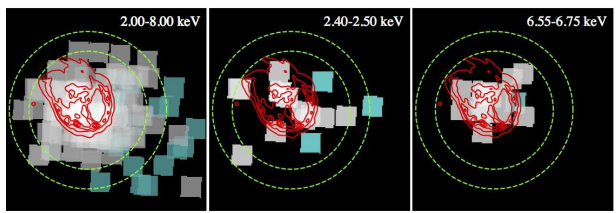}
     \end{center}
     \caption{Images of the SXS showing individual counts as a pixel on the
     sky.  Blue boxes are counts included after relaxing the angular
     distance criterion.  The red contours trace the Chandra emission, and
     the green circles show radii of 1$.\!\!^\prime$5 and 2$.\!\!^\prime$0
     from the Chandra peak.  The counts in Fe K (right) correspond well to
     the remnant extent, while some of the counts in the other bands are
     outside the bounds of the remnant.}
     \label{fig:sxscounts}
\end{figure}

We analyzed the cleaned event data of the final pipeline processing (Hitomi
software version 6, CALDB version 7) with the standard screening for both
SXS and SXI \citep{Angelini16}, with one exception.  To maximize the good
SXS observing time, we relaxed the requirement that eliminates data when
the aimpoint is further than 1$.\!\!^\prime$5 from the target position.
Using a maximum angular offset of 2$.\!\!^\prime$2 ensures that at least
50\% of the SNR is still in the FoV, and it increases the good SXS exposure
time from 2,610 s to 3,737 s (by 43\%) and the total SXS counts in the
2--10 keV band from 198 to 233 (by 18\%).  Relaxing this criterion
increased the counts in the Fe\,\textsc{xxv} He$\alpha$ band (defined in
section \ref{subsect:sxs_fe_analysis}) from 16 to 17, and in the
S\,\textsc{xv} He$\alpha$ band (defined in section
\ref{subsect:sxs_s_analysis}) from 13 to 16.  As we show in section
\ref{sect:sxs_analysis}, with the very low SXS background and very high
spectral resolution, this small number of counts is sufficient to derive
interesting constraints for the line centers.  Some of the additional
broad-band counts are from the region outside the N132D emission peak, as
shown in figure \ref{fig:sxscounts}, so they are likely background counts.
The extra counts in the lines are consistent with locations in the remnant,
also shown in figure \ref{fig:sxscounts}; in particular, to the extent that
we can infer locations from the $\sim$\,1\arcmin\ Hitomi PSF, the S counts
are found largely in the rim of the remnant, while the additional Fe K
count (and all the Fe K counts) are concentrated in the remnant center,
consistent with what is seen with XMM-Newton \citep{Behar01}.

We constructed an SXS source spectrum by extracting only GRADE Hp
(high-resolution primary) events from the entire SXS field of view of OBSID
100041010, and created the redistribution matrix file (RMF) with
\texttt{sxsrmf}, using the medium size option. The ancillary response file
(ARF) was generated with \texttt{aharfgen}, using a high-resolution Chandra
image as input to the ray-tracing.  A non-X-ray background (NXB) spectrum
with the same sampling of magnetic cut-off rigidity as the observation and
with identical filtering as the source data (except for Earth elevation
criteria) was extracted from the SXS archive NXB event file using
\texttt{sxsnxbgen}.  In the 2--10 keV band, we expect $23.2 \pm 0.6$ NXB
counts, about 10\% of the observed count rate, and corresponding to
$\sim$\,0.4 counts per spectral resolution element per 100 ks.  In the
narrow bands used for the analysis that follows, the NXB count rate is less
than 5\% of the observed rate as the SXS NXB is almost featureless and
nearly constant over the energy range (Eckart et al.~in prep.).

For the SXI, both OBSIDs 100041010 and 100041020 were used, although for
the former we enforced the requirement that the aimpoint be within
1$.\!\!^\prime$5 of the target to eliminate complications in constructing a
response for a source moving across the FoV.  For OBSID 100041020, we used
only times when the attitude was stable, although the source was not at the
expected aimpoint and was partially obstructed by the chip gaps (see figure
\ref{fig:att}).  The final good exposure time for the SXI was 35.4 ks.

An SXI spectrum was extracted from a 2$.\!^\prime$5 radius circle with
center (RA,Dec) = (\timeform{5h25m02.2s},\timeform{-69D38'39''}).  The NXB
spectrum was produced with \texttt{sxinxbgen}, using the entire SXI FoV
excluding the source in order to increase the statistics.  To properly
scale the NXB normalization between the full FoV and source region, the
instrumental lines of Au L$\alpha$ and L$\beta$ were used, producing a
scaling factor of 0.0070.  RMF and ARF files were generated with and
\texttt{sxirmf} and \texttt{aharfgen}, respectively.

\section{SXS Spectral Analysis}
\label{sect:sxs_analysis}

\begin{figure}[t]
     \begin{center}
          \FigureFile(8cm,8cm){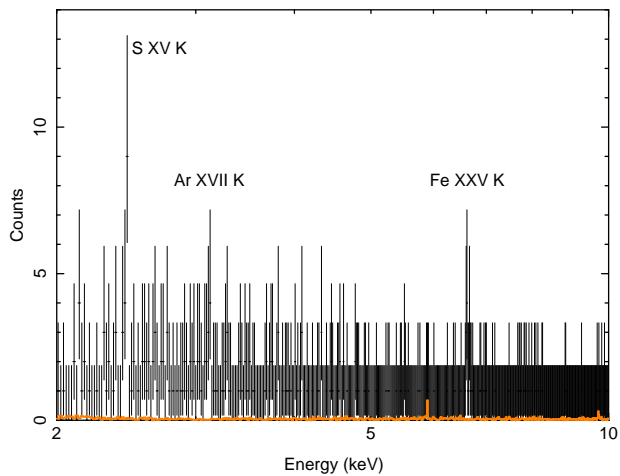}
     \end{center}
     \caption{Full-band SXS spectrum of N132D, showing counts with Poisson
     errorbars from \citet{Gehrels1986}. The orange points show the total
     estimated background, which has not been subtracted. Both spectra are
     binned to 16 eV for display purposes.}
     \label{fig:sxsfullspec}
\end{figure}

With only 233 counts, the SXS spectrum is dominated by Poisson low-count
statistics.  In addition, with the SXS gate valve closed, the bright
emission lines of C, O, Ne, and Mg below 2 keV are not observable.
However, three emission features are easily seen in the full-band spectrum
shown in figure \ref{fig:sxsfullspec}, the He$\alpha$ transition features
of He-like S ($\sim$\,2.45 keV), Ar ($\sim$\,3.1 keV), and Fe ($\sim$\,6.7
keV).  These lines are clearly detected in previous observations dating
back to BeppoSAX \citep{Favata97}, although the combination of an extended
source and lower sensitivity at these energies complicates their
measurement by X-ray grating instruments like Chandra/HETGS and
XMM-Newton/RGS.  From narrow bands centered on each expected line centroid,
the total number of counts and estimated NXB counts are 16 total
($0.30\pm0.07$ NXB) counts for S\,\textsc{xv} He$\alpha$; 14 total
($0.28\pm0.06$ NXB) counts for Ar\,\textsc{xvii} He$\alpha$; and 17 total
($0.8\pm0.1$ NXB) counts for Fe\,\textsc{xxv} He$\alpha$.  The
signal-to-noise of these features and the underlying continuum is
insufficient to obtain useful constraints on the metal abundance,
temperature, or velocity broadening of the emitting plasma.  However, as we
show below, given a reasonable spectral model from other sources, the
exquisite spectral resolution of SXS allows us to measure the line centers
and thus the average line-of-sight Doppler velocity of two of these
components, S and Fe.

All spectral fitting described below was performed with XSPEC v12.9.1d
\citep{Arnaud1996}, using atomic and non-equilibrium ionization (NEI)
emissivity data from AtomDB v3.0.8 \citep{Foster2012}, and abundance ratios
from \citet{AndersGrevesse1989}.  In each restricted
fitting region, we allowed only the line-of-sight velocity
and normalization of the appropriate thermal component (described below) to
vary in the initial fit.  While we include the cosmic X-ray background
(CXB), it is negligible; a reasonable model for the 2--10 keV contribution
of the CXB power law component with $\Gamma = 1.4$, $S$(2--10 keV) =
$5.4\times10^{-15}$ erg~cm$^{-2}$~s$^{-1}$~arcmin$^{-2}$ (e.g.,
\cite{Ueda1999,Bautzetal2009}) predicts a mean of 1.5 CXB counts across the
entire band and less than 0.1 CXB counts in any of the narrow spectral
analysis bands.  This is less than 1\% of the detected counts.  Galactic
foreground emission is negligible above 2 keV toward this direction ($l =
280\degree$, $b = -32.\!\!\degree8$).

\subsection{Iron Region Spectral Analysis}
\label{subsect:sxs_fe_analysis}

Fe K emission in N132D has been explored previously
\citep{Favata97,Behar01,XiaoChen08,Yamaguchi14b}, with the conclusion that
this feature is dominated by Fe\,\textsc{xxv} He$\alpha$ emission.  The
XMM-Newton/EPIC observations are successfully fit above 2.5 keV with a
two-temperature-component model with $kT =$ 0.89 and 6.2 keV
\citep{Behar01}.  The cooler component produces the strong soft emission
lines seen with XMM-Newton/RGS, and the hotter component explains the Fe K
emission.  In particular, \citet{Behar01} emphasize the lack of a
temperature component at $\sim$\,1.5 keV to explain the lack of observed
L-shell emission from Li-, Be-, and B-like Fe in the XMM-Newton spectrum.
A recent study using 240 ks of Suzaku data combined with a 60 ks NuSTAR
observation \citep{Bamba17} has produced a two-component broad-band
spectral model of N132D with a similar cool temperature ($kT \approx$ 0.8
keV) but that interprets the Fe K emission arising primarily from an
over-ionized, recombining plasma component with $kT_{\rm e} = 1.5$~keV,
$kT_{\rm init} > 20$~keV, and relaxation timescale $n_e
t~\approx~10^{12}$~s~cm$^{-3}$.  Crucially, the Suzaku data show a clear
detection of H-like Fe Ly$\alpha$ emission, indicating that an
under-ionized (ionizing) plasma is unlikely to contribute significantly to
the emission at these energies, and thus much of the otherwise unresolved
Fe K emission is likely due to He-like Fe rather than lower ionization
states.  

\begin{figure}[t]
     \begin{center}
          \FigureFile(8cm,8cm){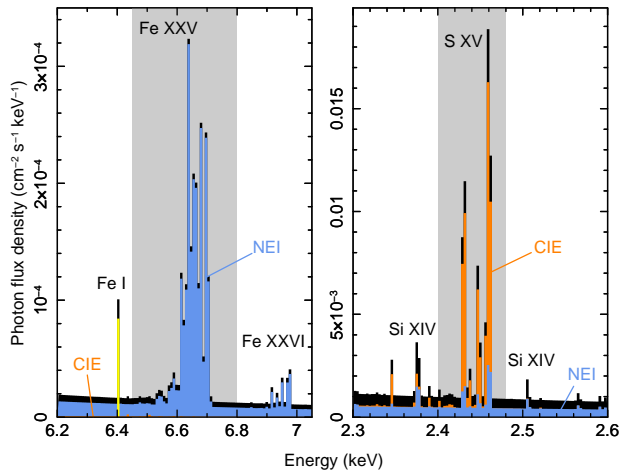}
     \end{center}
     \caption{N132D model spectrum, with contributions
     from individual components shown along with the total spectrum
     (black).  All components are plotted with zero
     velocity and no line broadening.  In the
     Fe\,\textsc{xxv} He$\alpha$ band (left), the NEI component (blue)
     dominates by a factor of $\sim$\,100 over the CIE component (orange).
     In the S\,\textsc{xv} He$\alpha$ band (right), the CIE component is
     $\sim 10$ times brighter than all other components.
     The gray shading indicates the bands used for spectral analysis; the S
     region is chosen to exclude contributions from Si\,\textsc{xiv}, while
     the Fe region is chosen to exclude the bright neutral Fe K line
     (yellow) and the Fe\,\textsc{xxvi} feature, but include possible
     contributions from lower ionization states of Fe near 6.5 keV.}
     \label{fig:model_spec}
\end{figure}

These previous observations provide confidence that we know where the line
centroid should be for the Fe K complex, and can cleanly measure the
line-of-sight velocity.  However, we emphasize that this is one possible
interpretation of a plasma with strong Fe\,\textsc{xxv} He$\alpha$ and
measurable Fe\,\textsc{xxvi} Ly$\alpha$ emission.  A more complicated
temperature structure, such as from multiple unassociated, spatially
unresolved components, could produce a very different complex of lines in
this spectral region. We address this possibility further in section
\ref{sect:disc}. To ease comparison to current work, we adopt the model
from \citet{Bamba17} as a baseline model, shown in figure
\ref{fig:model_spec} and table \ref{tab:sxs_params}.

\begin{table*}[p]
     \label{tab:sxs_params}
     \caption{Results of SXS Spectral Fitting.\footnotemark[$*$]}
     \begin{center}
          \small
          \begin{tabular}{lcccc}
               \hline
               Model Parameter                                                       & \multicolumn{2}{c}{Fe\,\textsc{xxv} fit}          & \multicolumn{2}{c}{S\,\textsc{xv} fit} \\ & no broadening                & with broadening$\dagger$                & no broadening                & with broadening$\dagger$ \\
               \hline
               \multicolumn{5}{l}{N132D CIE plasma (\texttt{vapec})} \\
               \hspace*{1em} $kT$ (keV)                                              & \multicolumn{4}{c}{\dotfill0.7\dotfill} \\
               \hspace*{1em} $Z_{\rm Si}$ (solar)                                    & \multicolumn{4}{c}{\dotfill0.6\dotfill} \\
               \hspace*{1em} $Z_{\rm S}$ (solar)                                     & \multicolumn{4}{c}{\dotfill0.9\dotfill} \\
               \hspace*{1em} $Z_{\rm Fe}$ (solar)                                    & \multicolumn{4}{c}{\dotfill0.4\dotfill} \\
               \hspace*{1em} $v_{\rm helio}$ (km~s$^{-1}$)                           & \multicolumn{2}{c}{\dotfill0\dotfill}             &  210$^{+370}_{-380}$    & 520$^{+770}_{-620}$ \\
               \hspace*{1em} $\sigma$ (km~s$^{-1}$)                                  & \multicolumn{2}{c}{\dotfill0\dotfill}             &  0                      & 520$^{+780}_{-340}$ \\
               \hspace*{1em} flux, 2--10 keV\footnotemark[$\ddagger$]         & \multicolumn{2}{c}{\dotfill8.8\dotfill}                  &  5.6$^{+2.9}_{-1.9}$    & 5.5$^{+3.1}_{-1.8}$ \\
               \hspace*{1em} flux, fitting band\footnotemark[$\ddagger$]      & \multicolumn{2}{c}{\dotfill0.006\dotfill}                &  1.3$^{+0.4}_{-0.2}$    & 1.3$^{+0.4}_{-0.2}$ \\
               \hline
               \multicolumn{5}{l}{N132D NEI plasma (\texttt{vrnei})} \\
               \hspace*{1em} $kT$ (keV)                                              & \multicolumn{4}{c}{\dotfill1.5\dotfill} \\
               \hspace*{1em} $kT_{\rm init}$ (keV)                                   & \multicolumn{4}{c}{\dotfill80\dotfill} \\
               \hspace*{1em} $n_et$ ($10^{12}$ s cm$^{-3}$)                          & \multicolumn{4}{c}{\dotfill0.9\dotfill} \\
               \hspace*{1em} $Z_{\rm Si}$ (solar)                                    & \multicolumn{4}{c}{\dotfill0.4\dotfill} \\
               \hspace*{1em} $Z_{\rm S}$ (solar)                                     & \multicolumn{4}{c}{\dotfill0.4\dotfill} \\
               \hspace*{1em} $Z_{\rm Fe}$ (solar)                                    & \multicolumn{4}{c}{\dotfill0.5\dotfill} \\
               \hspace*{1em} $v_{\rm helio}$ (km~s$^{-1}$)                           & 1440$^{+100}_{-1000}$  & 1140$^{+640}_{-810}$        & 1440 & 1140 \\
               \hspace*{1em} $\sigma$ (km~s$^{-1}$)                                  & 0                      & 510$^{+1060}_{-330}$        &  0   & 510 \\
               \hspace*{1em} flux, 2--10 keV\footnotemark[$\ddagger$]         & 9.5$^{+4.5}_{-3.0}$           & 9.7$^{+4.2}_{-3.2}$         & 6.1  & 6.2 \\
               \hspace*{1em} flux, fitting band\footnotemark[$\ddagger$]      & 0.48$^{+0.25}_{-0.16}$        & 0.49$^{+0.24}_{-0.16}$      & 0.34 & 0.34 \\
               \hline
               \multicolumn{5}{l}{CXB power law} \\
               \hspace*{1em} $\Gamma$                                                & \multicolumn{4}{c}{\dotfill1.54\dotfill} \\
               \hspace*{1em} flux, 2--10 keV\footnotemark[$\ddagger$]                 & \multicolumn{4}{c}{\dotfill0.040\dotfill} \\
               \hspace*{1em} flux, fitting band\footnotemark[$\ddagger$]              & \multicolumn{2}{c}{\dotfill0.0016\dotfill}      & \multicolumn{2}{c}{\dotfill0.0006\dotfill}  \\
               \hline
               spectral fitting band                                                 & \multicolumn{2}{c}{\dotfill6.45--6.80 keV\dotfill}             & \multicolumn{2}{c}{\dotfill2.40--2.48 keV\dotfill} \\
               C-stat / d.o.f.                                                       & 107.9 / 696         & 106.5 / 695        & 61.0 / 157   & 59.1 / 156 \\
               goodness-of-fit (KS)\footnotemark[$\S$]                               & 24\%                & 20\%               & 62\%         & 31\% \\
               goodness-of-fit (CvM)\footnotemark[$\S$]                              & 35\%                & 21\%               & 62\%         & 46\% \\
               \hline
               \multicolumn{5}{l}{\parbox{120mm}{
               \footnotesize
               \par \noindent
               \begin{tabnote}
                    \footnotemark[$*$]Unless noted otherwise, values
                    without quoted uncertainties are fixed. 
                    Uncertainties are 90\% confidence limits.\\
                    \footnotemark[$\dagger$]Results with broadening are
                    from inference with a Gaussian prior with
                    $\sigma = 1000$ km~s$^{-1}$.\\
                    \footnotemark[$\ddagger$]Flux is given in units of
                    $10^{-12}$ erg cm$^{-2}$ s$^{-1}$. The ratio of the
                    \texttt{vapec} and \texttt{vrnei} component
                    normalizations was fixed to that in \citet{Bamba17}. \\
                    \footnotemark[$\S$]``Goodness-of-fit'' is the
                    percentage of simulated observations with lower fit
                    statistic than the real data, as described in 
                    section \ref{subsect:sxs_fe_analysis}. 
               \end{tabnote}
               }}
          \end{tabular}
     \end{center}
     \vspace{-3mm}
\end{table*}

\begin{figure*}[t]
     \begin{center}
          \FigureFile(75mm,80mm){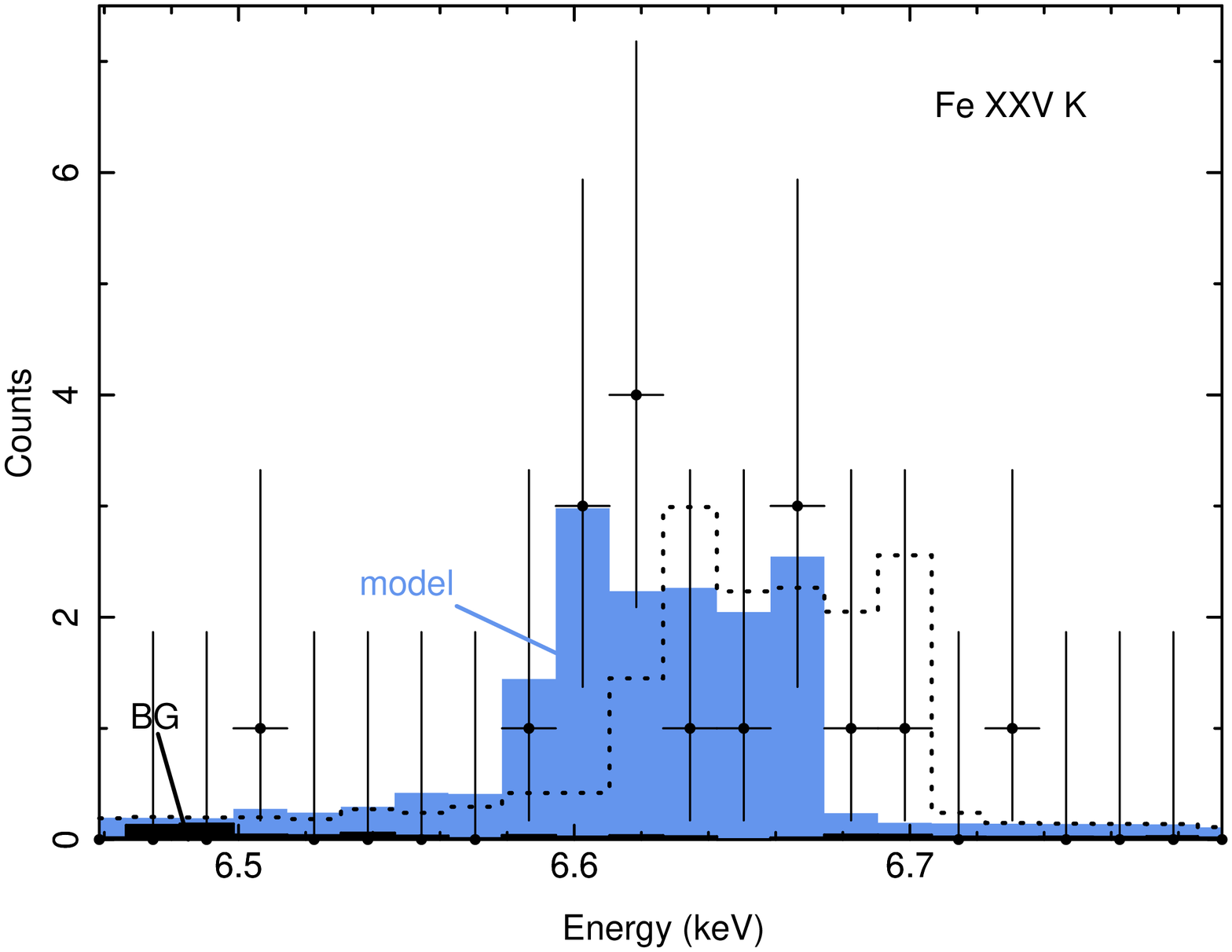}	
          \hspace*{10mm}
          \FigureFile(75mm,80mm){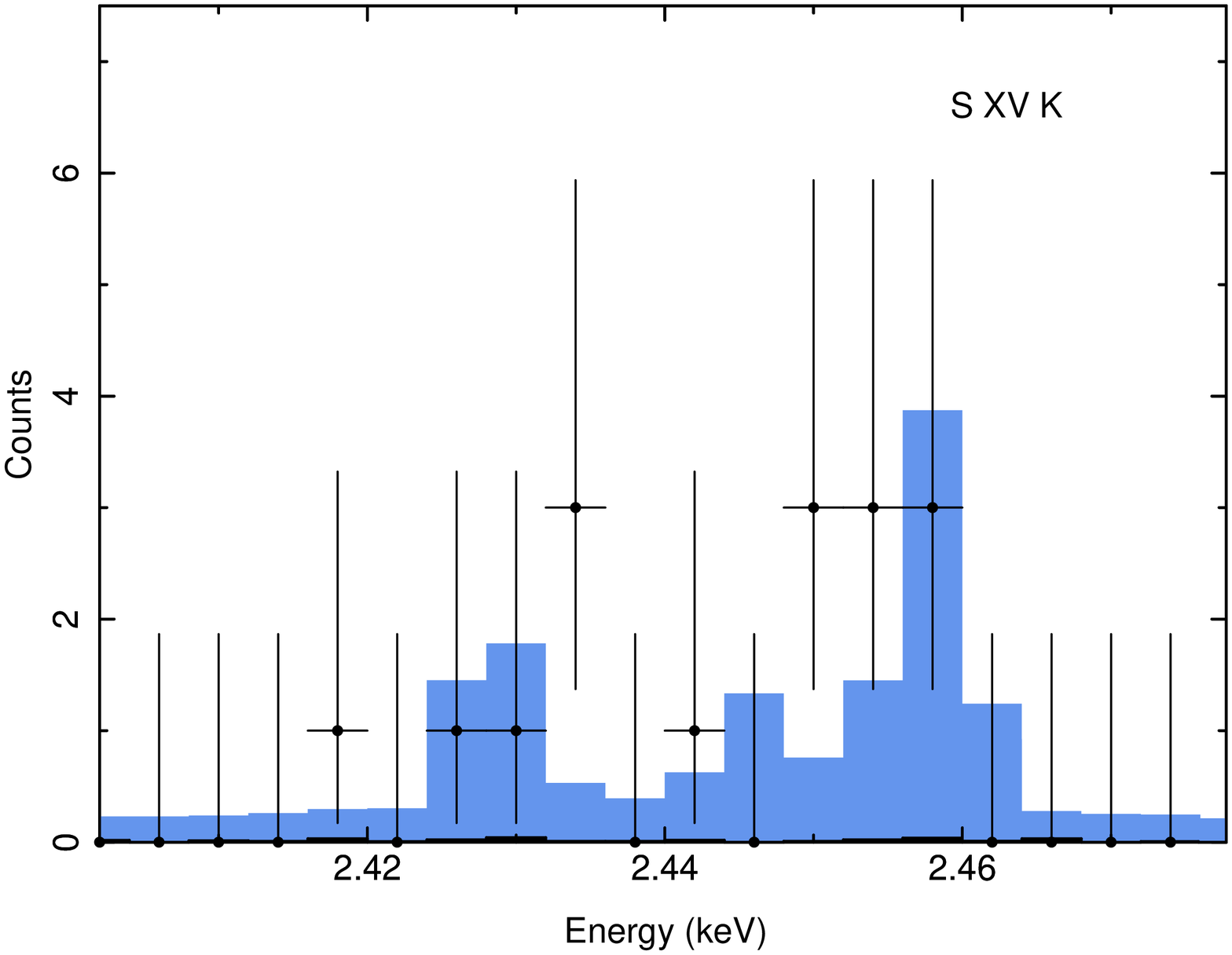}	
     \end{center}
     \caption{SXS spectra of the (left) Fe\,\textsc{xxv}
     He$\alpha$ and (right) S\,\textsc{xv} He$\alpha$ fitting regions.  The
     data points are detected SXS counts with Poisson error bars from
     \citet{Gehrels1986}.  In both panels, the blue shaded region shows the
     best-fit model, and the black shaded region, barely visible, shows the
     estimated total background.  In the left panel, the dotted line shows
     the model with velocity fixed at $v_{\rm helio,LMC}
     =$ 275 km~s$^{-1}$.  The Fe spectrum is binned to 16 eV and S binned
     to 4 eV for display purposes.}
     \label{fig:fe_spec}
     \label{fig:s_spec}
\end{figure*}

The Fe\,\textsc{xxv} He$\alpha$ complex, shown in figure \ref{fig:fe_spec},
was fit within the energy range 6.45--6.80 keV.  This range includes
sufficient width to constrain the continuum and measure velocity shifts up
to $\sim$\,7000~km~s$^{-1}$, but avoids contamination from a possible 6.4
keV Fe K line and any H-like Fe features.  It is clear from figure
\ref{fig:model_spec} that in this very clean fitting region the model is
dominated by emission from the recombining plasma component by at least a
factor of 100 over the cooler collisional ionization equilibrium (CIE)
component.  Therefore, while we included the entire model with all
components for the Fe region fit, we only allowed parameters related to the
NEI component to vary.  To allow for differences in the observed flux due
to the smaller SXS FoV and attitude drift, we fixed the ratio of the CIE to
NEI component normalizations to that derived by \citet{Bamba17}, and
allowed the NEI flux to vary along with the line-of-sight velocity.  The
CIE component was modeled by a variable-abundance \texttt{vapec} model in
XSPEC, while the NEI component was modeled by a variable-abundance
recombining plasma model, \texttt{vrnei}.  We included a single Gaussian
broadening parameter to allow for thermal and turbulent broadening as well
as unresolved bulk motion.  

\begin{figure*}[p]
     \begin{center}
          \FigureFile(80mm,80mm){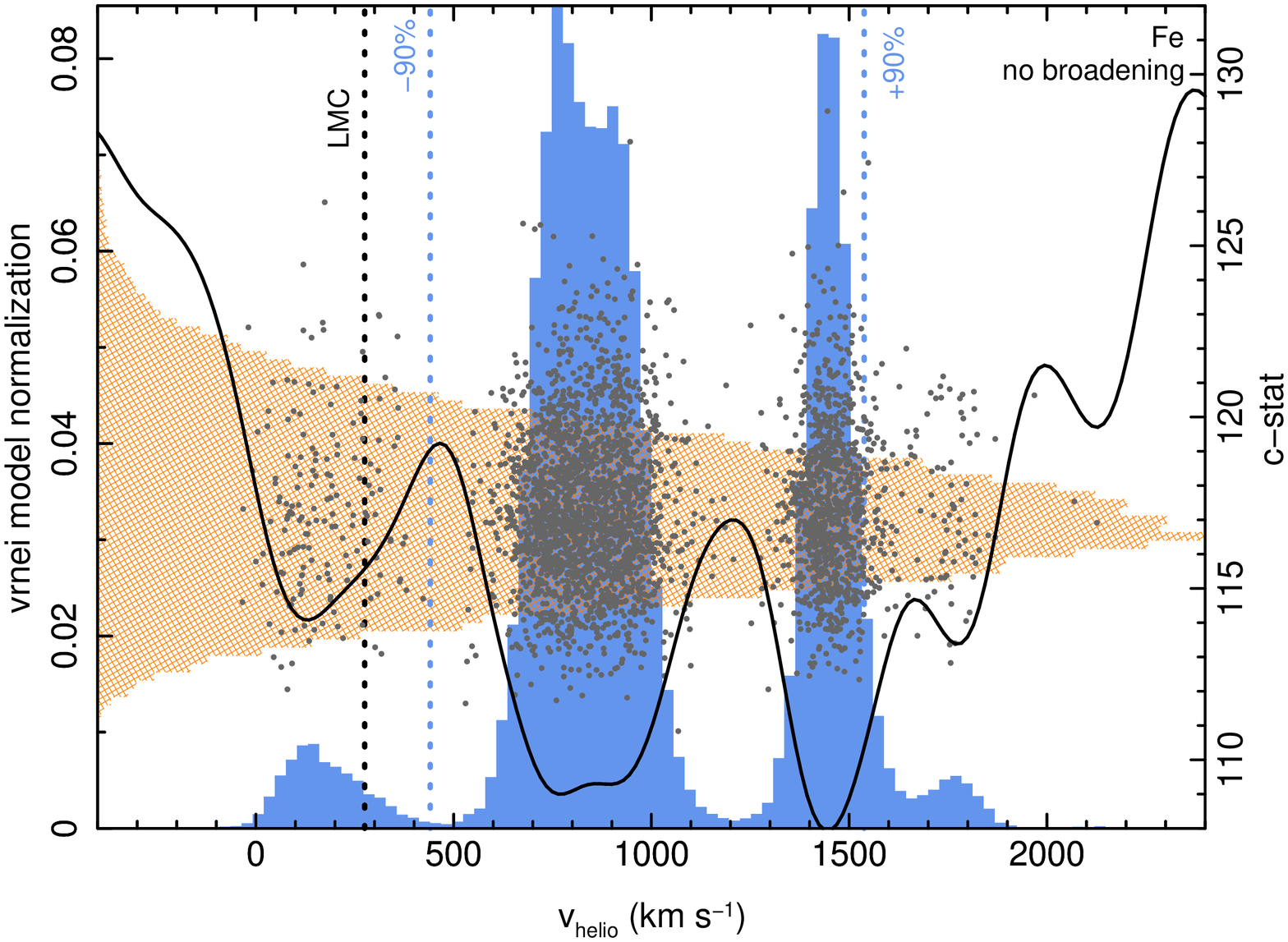}
          \FigureFile(80mm,80mm){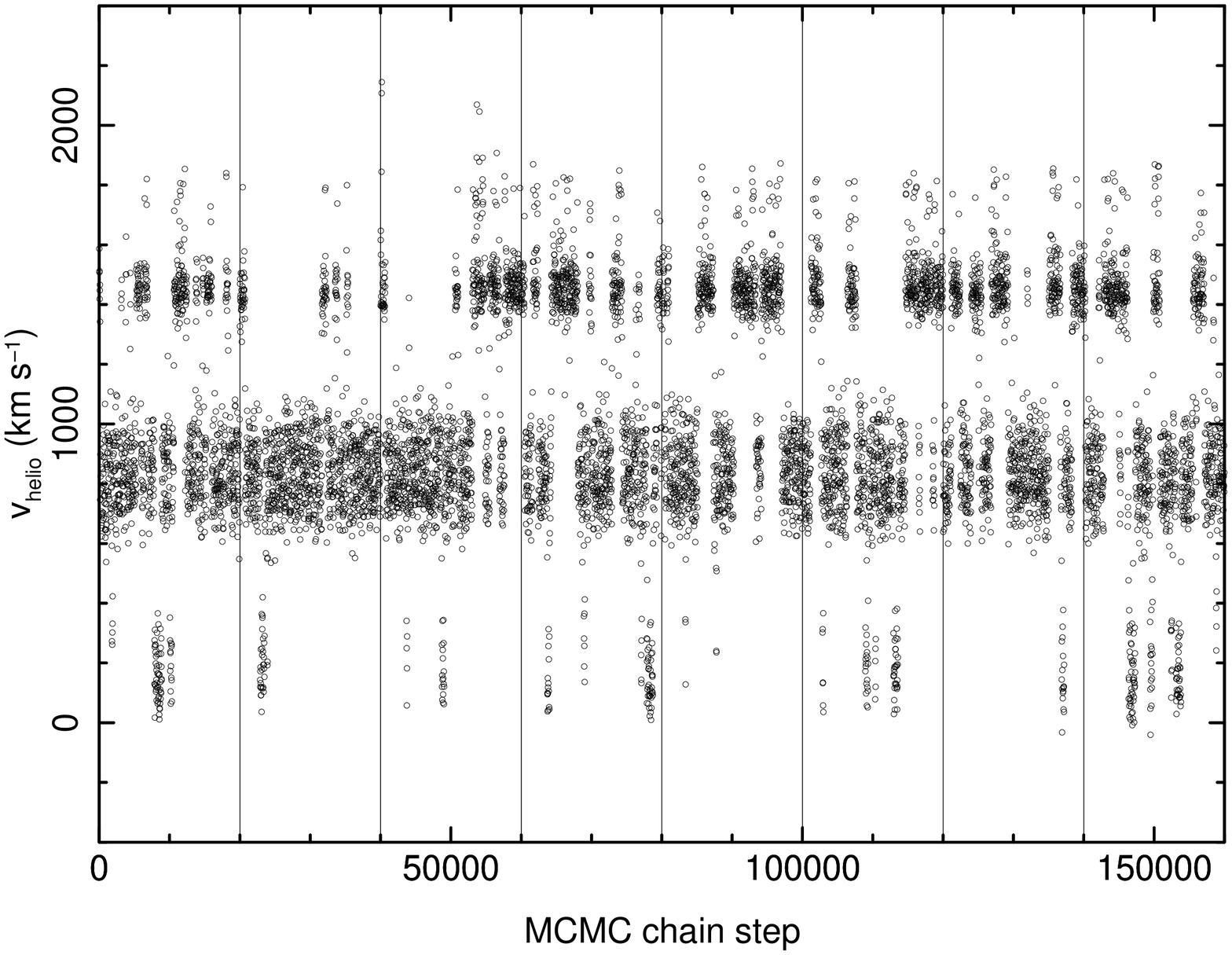}
     \end{center}
     \caption{(left) Posterior probability distributions
     of the \texttt{vrnei} model normalization (orange) and velocity (blue)
     from the Fe K region fitting without broadening, calculated from the
     MCMC analysis as described in the text.  Points show sample MCMC
     chain steps, indicating that there is
     no correlation between the two parameters.  The black line shows the
     \texttt{cstat} value from fit statistic minimization, as a function of
     best-fit velocity.  One peak of the MCMC velocity
     distribution coincides with the best-fit velocity distribution, and
     other local peaks coincide with local \texttt{cstat} minima,
     indicating both maximum likelihood methods produce the same result.
     The dotted lines delineate the central 90\% credible interval and note
     the local LMC velocity.  (right) MCMC chain values for
     the Fe K velocity plotted against chain step, showing that the
     long-term variations of each chain are well-behaved and the posterior
     distribution is well-sampled.  Vertical lines differentiate the eight
     individual 20,000-step simulation chains. Steps within chains are in
     time order with one out of every ten steps shown for clarity.}
     \label{fig:fe_post_nobroad}
\end{figure*}

\begin{figure*}[p]
     \begin{center}
          \FigureFile(80mm,80mm){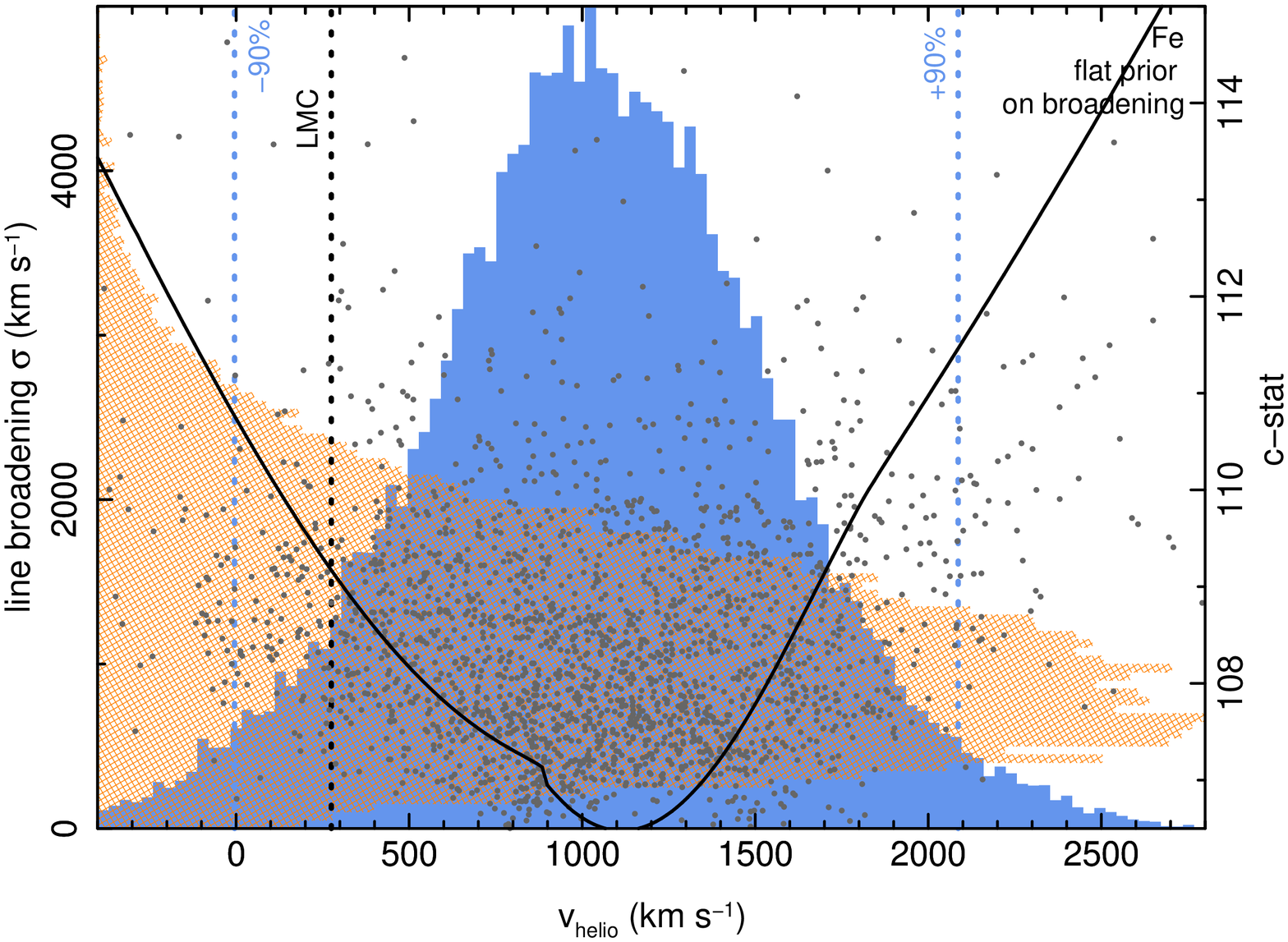}
          \FigureFile(80mm,80mm){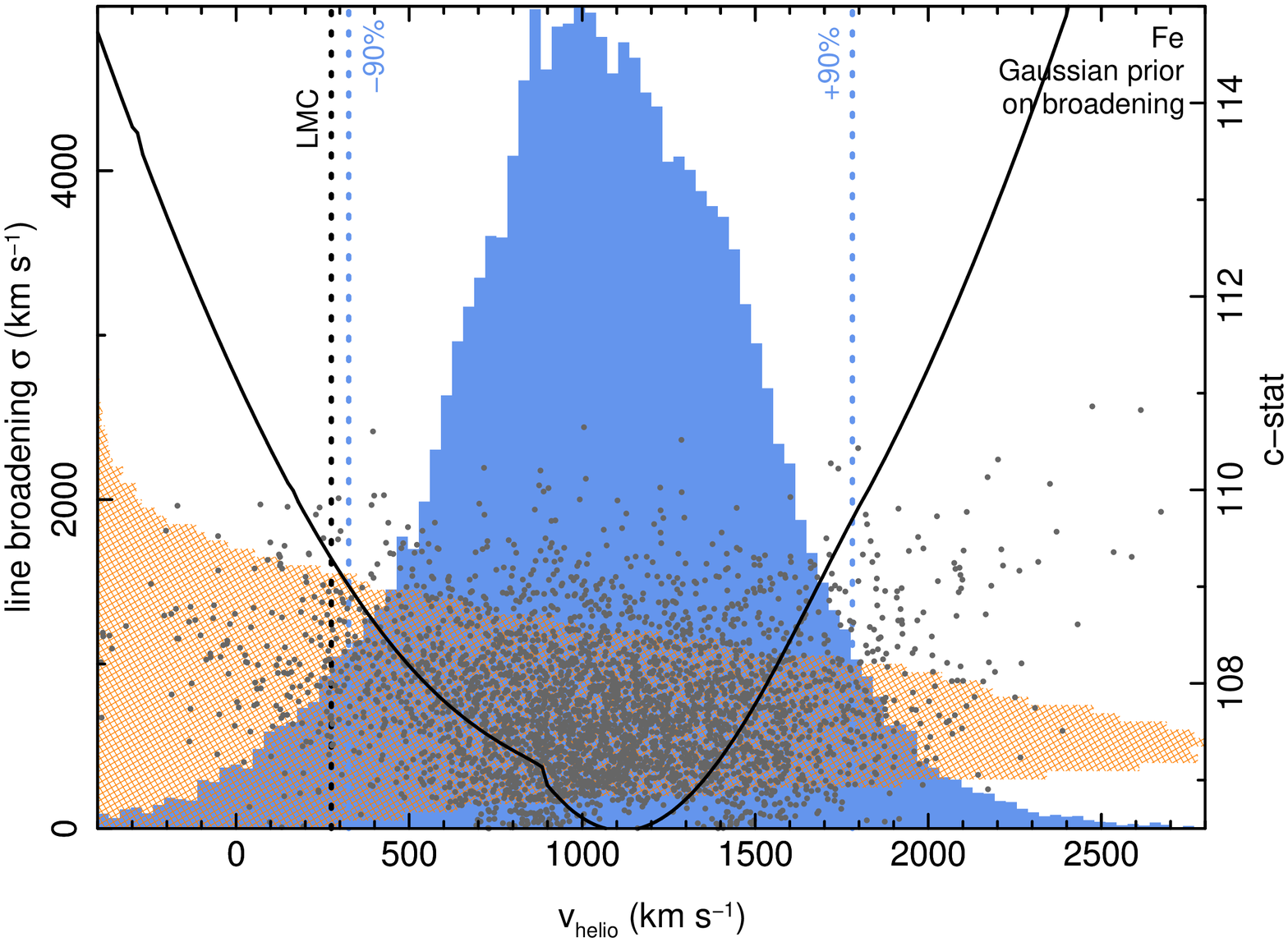}
     \end{center}
     \caption{Posterior probability distributions of the \texttt{vrnei}
     model broadening width (orange) and velocity (blue) from the Fe K
     region fitting including line broadening.  Notations are the same as
     in figure \ref{fig:fe_post_nobroad}.  The left panel shows results
     with flat prior on the line width, while the right
     panel shows results imposing a Gaussian prior with
     1-$\sigma$ width of 1000 km~s$^{-1}$.  Both velocity distributions
     trace the \texttt{cstat} minimization well.  The flat prior produces a
     broader posterior distribution.}
     \label{fig:fe_post_broad}
\end{figure*}

Parameter estimation was performed in two ways.  First, maximum likelihood
estimation was done by minimizing the fit statistic, using \texttt{cstat}
in XSPEC, a modified \citet{Cash1979} statistic.  With the broadening width
fixed at zero, this fitting revealed a highly non-monotonic parameter space
for the velocity (see figure \ref{fig:fe_post_nobroad}), likely due to the
combination of low-count Poisson statistics in the data and discrete
spectral features in the model.  The best-fit velocity of $v_{\rm helio} =
1440$ km~s$^{-1}$ is significantly larger than the value of the local LMC
ISM surrounding N132D, $v_{\rm helio,LMC} = 275\pm4$ km~s$^{-1}$
\citep{VogtDopita2011}.  Allowing a free broadening width eliminated this
non-monotonicity (see figure \ref{fig:fe_post_broad}), resulting in a
best-fit $v_{\rm helio} = 1140$ km~s$^{-1}$ and broadening of $\sigma =
510$ km~s$^{-1}$.

Second, to fully explore parameter space, we performed Markov chain Monte
Carlo (MCMC) simulations within XSPEC using Bayesian inference.  
These simulations were run with and without velocity broadening, using both
a flat (uniform) prior distribution and a Gaussian prior distribution for
the broadening width.  The width of the Gaussian prior distribution was
chosen to reflect current upper limits on the velocity broadening.  In
particular, observations with CCD-based X-ray observatories such as Suzaku
(e.g.  \cite{Bamba17}) have not found measurable broadening.  The typical
spectral resolution of such instruments near 6 keV is $\sim$\,150--180 eV
FWHM, depending on the epoch of observation, with a typical 1-$\sigma$
calibration uncertainty of 5\%\footnote{See Table 3.2 and Figure 7.11 of
the Suzaku Technical Description,
ftp://legacy.gsfc.nasa.gov/suzaku/nra\_info/suzaku\_td\_xisfinal.pdf.}.
This calibration uncertainty can be thought of as an upper limit on the
detectable line broadening velocity.  Since the broadening is a
convolution, this extra velocity component adds in quadrature with the
instrumental width.  We find that a 5\% increase on the 150--180 eV FWHM
instrumental width is equivalent to an extra broadening component with FWHM
of 48--58 eV, or $\sigma$ = 900--1100~km~s$^{-1}$ in the center of our
fitting band.  We therefore adopted 1000~km~s$^{-1}$ as a natural
1-$\sigma$ width to use for the Gaussian prior distribution.  We performed
MCMC simulations using both the flat, uninformative prior and the weakly
informative Gaussian prior.

The MCMC results are consistent with the local \texttt{cstat} minima in
velocity parameter space for fits with and without broadening, as shown by
the MCMC posterior probability distributions in figures
\ref{fig:fe_post_nobroad} and \ref{fig:fe_post_broad}.  In particular, the
complicated velocity posterior distribution shows up clearly in the MCMC
runs without broadening, but with the most likely value (highest mode) near
$v_{\rm helio} = 800$ km~s$^{-1}$ instead of 1400 km~s$^{-1}$ as found in
the \texttt{cstat} minimization.  The MCMC chain steps shown in figure
\ref{fig:fe_post_nobroad} (right) indicate that the simulation is
well-behaved and samples the posterior distribution adequately despite the
multimodal structure. The runs with broadening result in Gaussian posterior
distributions with peak near 1000 km~s$^{-1}$.  Using either a Gaussian or
Cauchy form for the chain proposal distribution produced the same results.

We used these posterior distributions to obtain central credible intervals
on $v_{\rm helio}$.  For the fit with no broadening, a single interval is
uninformative due to the complicated structure.  We obtain a 68\% credible
interval of 730--1460 km~s$^{-1}$, 90\% interval of 440--1540 km~s$^{-1}$,
and 95\% interval of 160--1620 km~s$^{-1}$.  A line-of-sight velocity
consistent with $v_{\rm helio,LMC}$ is ruled out at 93\% confidence under
this model.  With broadening, a single credible interval is sufficient to
characterize the Gaussian-shaped distribution, and we find 90\% credible
intervals of 330--1780 km~s$^{-1}$ for broadening with a Gaussian prior
distribution, and 0--2090 km~s$^{-1}$ for a flat prior. The conservative
gain uncertainty of $\pm$\,2~eV (see section \ref{sect:obs}) produces a
systematic uncertainty of $\pm$\,90~km~s$^{-1}$, well within the
statistical uncertainty.  It is apparent that imposing an flat,
uninformative prior on the broadening width distribution allows unrealistic
values exceeding $\sigma = 3000$ km~s$^{-1}$ with a broad tail to very high
values.  This greatly exceeds the thermal width of an Fe emission feature
at 2 keV ($\sigma \sim 50$ km~s$^{-1}$), and requires either extreme
turbulence or very large bulk motions.  If we adopt the results with the
Gaussian prior, which has sufficient width to allow a blueshifted and
redshifted component separated by up to $\sim$\,2000 km~s$^{-1}$, a mean
line-of-sight velocity consistent with $v_{\rm helio,LMC}$ is ruled out at
91\% confidence under this model.  The model parameters are listed in table
\ref{tab:sxs_params}.

\clearpage
\clearpage

The measured photon flux in the fitting band,
$4.6^{+2.3}_{-1.4}\times10^{-5}$~ph~cm$^{-2}$~s$^{-1}$, is more than a
factor of two higher than previous estimates of the Fe~K$\alpha$ line flux,
e.g.  $1.83 \pm 0.17 \times10^{-5}$~ph~cm$^{-2}$~s$^{-1}$
(\cite{Yamaguchi14b}; errors are 90\%).  This is likely due to a
combination of the Hitomi attitude uncertainty and the use of a broad-band
X-ray image to produce the response files.  While much of this broad-band
X-ray emission is found in a shell with diameter $\sim\,$2\arcmin, the
Fe~K$\alpha$ emission appears centrally concentrated (e.g.,
\cite{Behar01}).  Using the more spatially extended broad-band image
produces a lower response as some of the PSF-broadened flux falls outside
of the $3\arcmin\times3\arcmin$ SXS FOV, thereby increasing the inferred
model flux for a given count rate.  Our inclusion of data with large
pointing offset of up to 2.2$\arcmin$ and the large attitude drift
undoubtedly exacerbate this effect.  For this reason, the flux calibration
is so uncertain that a flat, uninformative prior is a good representation
of our knowledge of the SXS effective area for this observation.

Once the minimum fit statistic and parameter distribution function were
determined, we explored the effects of adjusting other \texttt{vrnei}
parameters within a reasonable range of uncertainty.  In addition, we ran
fits testing plasma models with higher over-ionization (setting $n_et$ to a
small value), under-ionization (an ionizing plasma, setting $kT_{\rm init}
< kT$), and collisional ionization equilibrium (CIE, setting $kT_{\rm init}
= kT$).  The fit statistic was consistent in all cases, indicating that we
cannot distinguish between various ionization states with the Hitomi/SXS
data alone.  In all cases, neither the best-fit velocity nor its posterior
distribution from the MCMC analysis changed appreciably, indicating that
our results are insensitive to the exact emission model used so long as it
is not highly complex.

Neither the XSPEC \texttt{cstat} statistic nor the MCMC analysis provides
an estimate of the goodness of fit.  We used two tests available in XSPEC,
Kolmogorov-Smirnov (KS) and Cramer-von Mises (CvM), both of which treat the
observed and model spectra as empirical distribution functions and compute
a statistical difference between the two.  Drawing parameter values for
velocity, normalization, and broadening width from the full posterior
distributions, we performed 1000 simulations of the observed 3.7 ksec
spectrum for the fits with and without broadening.  These simulated spectra
were then fit with the model, and the resulting KS and CvM test statistics
were compared with the values from the original fits.  For the fit without
broadening, 24\% of the realizations produced a smaller KS statistic than
the best fit, and 35\% produced a smaller CvM statistic.  For the fits with
broadening, the fractions were 20\% for KS and 21\% for CvM.  We can only
say that our best-fit models are not statistically inconsistent with the
data.

Since this asymmetric velocity structure is unexpected, we constrained a
potential blue-shifted emission feature by adding a second \texttt{vrnei}
component with identical model parameters.  The velocity of the first
component was fixed to the best-fit value of 1140 km~s$^{-1}$, while that
of the new component was fixed to $-590$ km~s$^{-1}$, to force symmetry
about $v_{\rm helio,LMC}$.  The \texttt{vrnei} normalizations, initially
equal, were allowed to vary independently.  We find that a blue-shifted
feature is allowed at up to 30\% of the flux of the redshifted component,
with a similar fit statistic and goodness-of-fit measure.  Varying the
blueshift within a reasonable range did not improve the fit or change the
upper limit to its flux.  The best-fit broadening width ($\sigma \sim$ 500
km~s$^{-1}$ or FWHM $\sim$ 1200 km~s$^{-1}$) allows some blueshifted
component, but the emission-weighted mean velocity is not centered on the
LMC velocity. We conclude that the bulk of the He-like-iron-bearing
material is receding asymmetrically, at a velocity $\sim$\,800 km~s$^{-1}$
with respect to the swept-up ISM surrounding N132D.

\subsection{Sulfur Region Spectral Analysis}
\label{subsect:sxs_s_analysis}

\begin{figure*}[t]
     \begin{center}
          \FigureFile(80mm,80mm){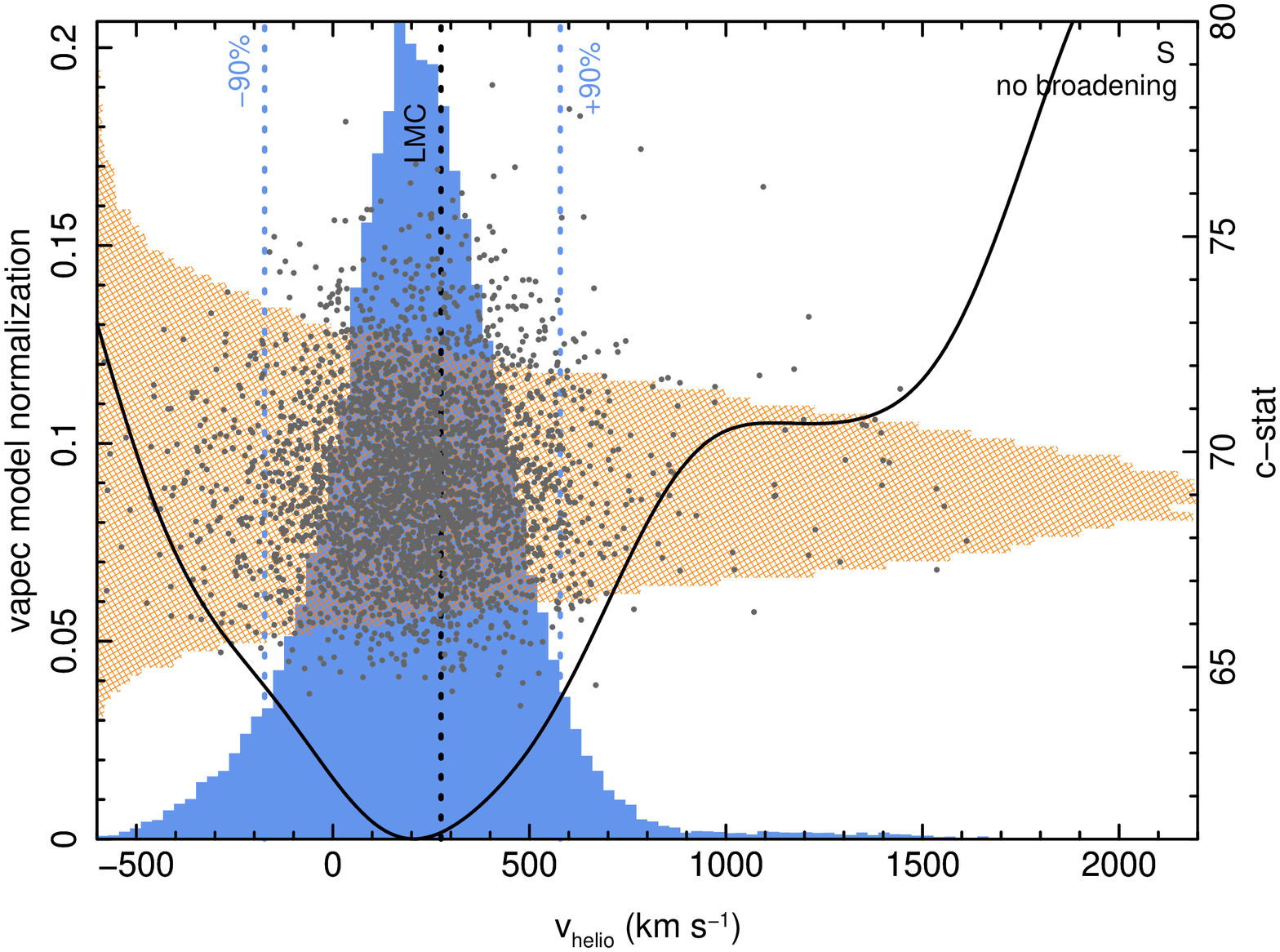}
          \FigureFile(80mm,80mm){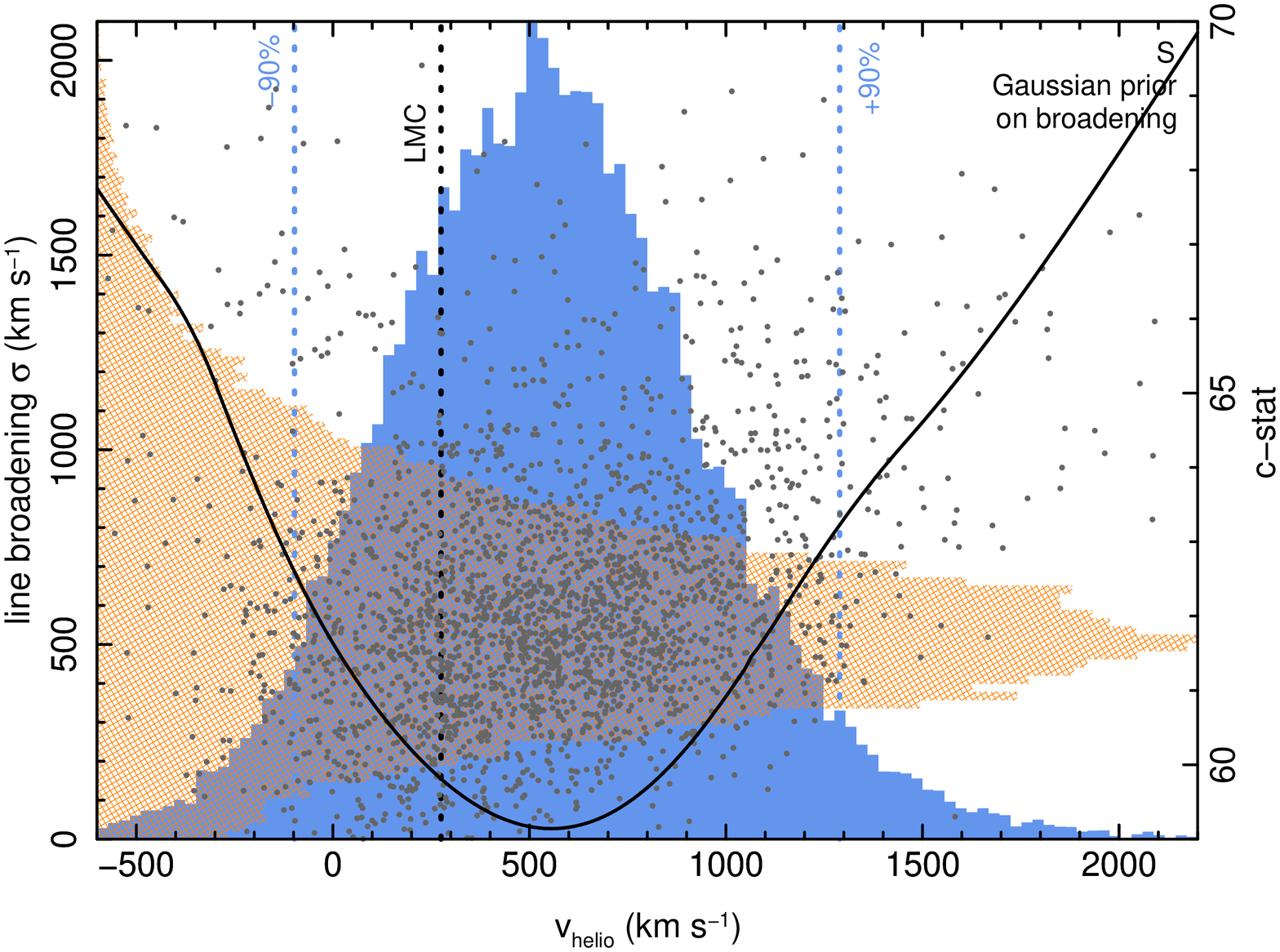}
     \end{center}
     \caption{Posterior probability distributions for the S region fit.  In
     the left panel, \texttt{vapec} model broadening is fixed at zero, and
     the posterior model normalization is shown in orange.
     In the right panel, a Gaussian prior with width
     $\sigma = 1000$ km~s$^{-1}$ is imposed on the broadening, and the
     posterior broadening distribution is shown in orange.  The velocity
     posterior distribution is shown in blue in both panels.  Other
     notations are the same as in figure \ref{fig:fe_post_nobroad}.  Both
     velocity posterior distributions are Gaussian in shape and trace the
     \texttt{cstat} minimization well.  The model with broadening produces
     a broader distribution shifted to higher velocity, but still
     consistent with the local LMC velocity.}
     \label{fig:s_post}
\end{figure*}

Spectral fitting of the S\,\textsc{xv} He$\alpha$ line proceeded in a
similar manner to the Fe K region.  We restricted the energy range to
2.40--2.48 keV, leaving 16 total counts of which $0.30\pm0.07$
($\sim$\,2\%) are estimated to be from the NXB.  Consistent with other
recent work, we interpret the S\,\textsc{xv} He$\alpha$ emission to arise
predominantly from a CIE plasma with $kT \sim$ 1 keV
\citep{Behar01,Borkowski07,XiaoChen08}.  In our baseline model, the CIE
component dominates the NEI emission by a factor of $\sim$\,5--10 in this
region.  Thus we allowed some small contamination from the high-redshift
NEI emission by freezing the velocity and broadening of the \texttt{vrnei}
component to the best-fit values, and fixed the ratio of the \texttt{vapec}
to \texttt{vrnei} normalizations to that found by \citet{Bamba17}.  Only
the velocity and normalization of the CIE \texttt{vapec} component were
allowed to vary in the initial fit, but as with the Fe fit, we included
broadening with similar priors to explore the effect on the derived
velocity.  The S region spectrum and model are shown in figure
\ref{fig:s_spec}, posterior probability distributions are shown in figure
\ref{fig:s_post}, and best-fit parameters are given in table
\ref{tab:sxs_params}.

Using the \texttt{cstat} maximum likelihood estimator, we obtain a best-fit
line-of-sight velocity of $v_{\rm helio} = 210$~km~s$^{-1}$ with broadening
fixed at zero.  Allowing a single broadening component results in $v_{\rm
helio} = 520$~km~s$^{-1}$ with $\sigma = 520$ km~s$^{-1}$.  As with the Fe
fitting, the posterior distributions in figure \ref{fig:s_post} are
considerably wider when broadening is included, with 90\% credible
intervals on $v_{\rm helio}$ of $-$170 to $+$580~km~s$^{-1}$ with no
broadening and $-$100 to $+$1290~km~s$^{-1}$ with a
Gaussian prior on broadening with $\sigma =
1000$~km~s$^{-1}$.  Unlike for Fe, the velocity of the S component is
completely unconstrained with a flat broadening prior.
Our adopted SXS gain uncertainty of $\pm$\,2~eV (245~km~s$^{-1}$; see
section \ref{sect:obs}) is again well within this statistical uncertainty,
which itself is consistent with the local LMC velocity of 275 km~s$^{-1}$.  

We performed additional spectral fitting, allowing $kT$ of the CIE
component and $kT$, $kT_{\rm init}$, $n_et$, and $\sigma$ of the
recombining plasma component to vary over a broad range as in the Fe region
fitting described in the previous section.  The best-fit velocity and
credible intervals did not change.  We performed the same goodness-of-fit
tests to the S region fits as the Fe region fits, finding that 30--60\% of
the simulated datasets produced a smaller test statistic.  The model is
thus consistent with the data, and we conclude that the
He-like-sulfur-bearing gas is consistent with being at rest relative to the
local LMC ISM, if we assume that line broadening is small.

\subsection{Argon Region Spectral Analysis}
\label{subsect:sxs_ar_analysis}

Spectral fitting of the Ar\,\textsc{xvii} He$\alpha$ line is complicated by
both the low number of total counts (14) and the estimated contributions
from both CIE and NEI components.  In fact, the Ar abundance is not
constrained in either component, leading to a degeneracy between the
normalization and abundance in each component and further difficulty
fitting different velocities.  As a simple test, we fixed the
\texttt{vapec} and \texttt{vrnei} normalizations to the \citet{Bamba17}
values, fixed the Ar abundance to solar for both components, and fit a
single line-of-sight velocity and normalization.  The best-fit velocity is
$v_{\rm helio} = 2400$~km~s$^{-1}$, with a 90\% credible interval of
570--5900 km~s$^{-1}$.  This is consistent with both velocity ranges of
Fe\,\textsc{xxv} and S\,\textsc{xv}.  If the velocities are tied at the
offset to the best-fit values so that $v_{\texttt{vrnei}} =
v_{\texttt{vapec}} + 1200$~km~s$^{-1}$, the fit statistic is only slightly
worse (\texttt{cstat} = 81.2 vs.~80.8), and the best-fit values are $v_{\rm
helio} = 1800$ km~s$^{-1}$ for the \texttt{vrnei} component and 600
km~s$^{-1}$ for the \texttt{vapec}, with similar uncertainties.  Given the
uncertainties in the model, we can only conclude that the Ar\,\textsc{xvii}
fit is consistent with the Fe and S line results.

\section{SXI Spectral Analysis}
\label{sect:sxi_analysis}

\begin{figure}[t]
     \begin{center}
          \FigureFile(7cm,8cm){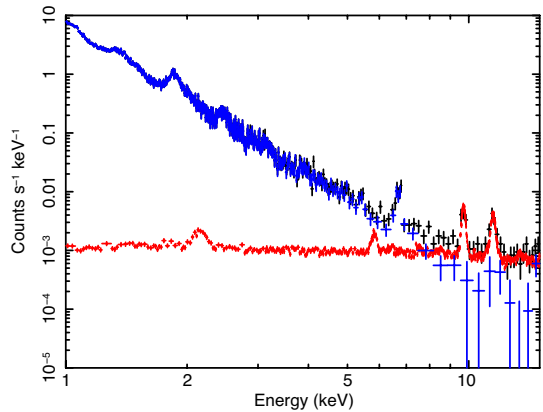}	
     \end{center}
     \caption{SXI spectrum of N132D.  Black data points show the full
     spectrum, red show the scaled NXB spectrum, and blue show the
     NXB-subtracted spectrum.  Emission over the background is clearly seen
     above 10 keV.}
     \label{fig:sxi_spec}
\end{figure}

For the following analysis, the same version of XSPEC, AtomDB, NEI
emissivity data, and abundance tables were used as in the analysis of the
SXS spectrum (see section \ref{sect:sxs_analysis}).  The NXB-subtracted
spectrum is shown in figure~\ref{fig:sxi_spec}. In the N132D observation,
the event and split thresholds are 600~eV and 30~eV, respectively.  Since
charge from a detected X-rays may be split among multiple CCD pixels, the
quantum efficiency (QE) can be affected by split events well above the
event threshold.  Given the limited amount of calibration information
available in these early observations, we conservatively exclude the energy
band below 2 keV in this study. 

We detect emission lines at $2.456\pm0.010$~keV and $6.68\pm0.04$~keV,
which correspond to the same He$\alpha$ lines of S and Fe detected in SXS,
respectively.  The SXI is affected by light leak when the satellite is in
daylight, which can result in an observed line center shift (Nakajima et
al.~in prep.).  We investigated the line center shift in the N132D data,
and confirmed that daylight illumination of the spacecraft has no effect.
The S\,\textsc{xv} He$\alpha$ line center is fully consistent with the
centroid of the line complex measured with SXS (see figure
\ref{fig:fe_spec}).  The Fe\,\textsc{xxv} He$\alpha$ line center is
marginally consistent with SXS within the uncertainty (see figure
\ref{fig:s_spec}), and likely includes some unresolved contribution from
Fe\,\textsc{xxvi} Ly$\alpha$ at $\sim$\,7 keV (see figure
\ref{fig:model_spec}).

\begin{figure}[t]
     \begin{center}
          \FigureFile(7cm,8cm){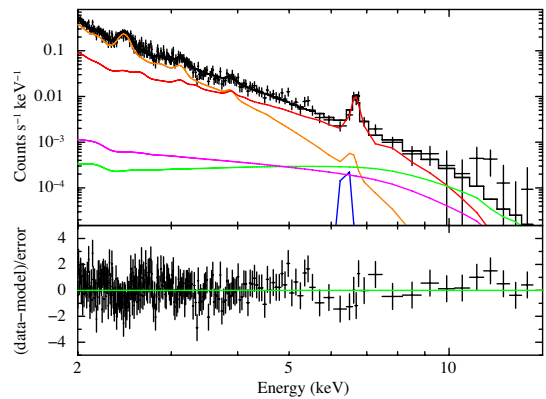}	
     \end{center}
     \caption{SXI spectrum of N132D fitted with the model discussed in the
     text, with individual components shown.  Two thin-thermal plasma
     components are shown with orange (\texttt{vapec}) and red
     (\texttt{vrnei}) lines.  The 6.4 keV Fe K line is shown in blue, the
     CXB power law component in green, and the marginally detected
     non-thermal component in magenta.}
     \label{fig:sxi_spec_fit}
\end{figure}

Following the SXS analysis, we adopted a spectral model with two
thin-thermal plasmas, a low-temperature \texttt{vapec} and high-temperature
\texttt{vrnei}.  From the model of \citet{Bamba17}, we also include a
6.4~keV neutral Fe K line, a non-thermal component, and the CXB.  In the
SXI analysis, the normalizations of the two plasmas are set to be free and
all the other thermal parameters are fixed to those of \citet{Bamba17}.
The normalization of the Fe~\textsc{i} K line was tied to that of the
\texttt{vrnei} component using the ratio of normalizations from
\citet{Bamba17}.  A power-law model was added for the possible non-thermal
component, with both photon-index $\Gamma$ and normalization allowed to
vary.  For the CXB, another power-law model with fixed parameters of
$\Gamma=1.4$ and surface brightness
$5.4\times10^{-15}$~erg~cm$^{-2}$~s$^{-1}$~arcmin$^{-2}$ in the 2--10~keV
band was used \citep{Ueda1999,Bautzetal2009}.  This CXB intensity is
expected from observations with previous X-ray imaging instruments with
similar PSF, and thus similar confusion limits.  Since we are in the
high-counts regime with at least 30 counts per spectral bin in the total
(unsubtracted) source spectrum and high statistics in the NXB spectrum, we
expect the background-subtracted spectral bins to be Gaussian distributed
and use $\chi^2$ minimization. We obtain $\chi^2 / $d.o.f.$=234/243$ and an
acceptable fit at the 90\% confidence level.  The best-fit model with
individual components is shown in figure~\ref{fig:sxi_spec_fit}.  To check
for potential bias in the use of $\chi^2$ statistics, we perform the fit
again excluding the poorest statistical region above 9 keV, and obtain
similar results.

The lower and higher temperature plasmas produce the majority of the
He$\alpha$ lines of S and Fe, respectively, consistent with the result of
the previous study (see also figure \ref{fig:model_spec}).  The best-fit
\texttt{vapec} normalization was $0.92\pm0.03$ of the value from
\citet{Bamba17}, while the \texttt{vrnei} was $0.86\pm0.10$ of their
best-fit value.  The model fitting results in a non-thermal component with
flux $1.3\pm1.1\times10^{-13}$~erg~cm$^{-2}$~s$^{-1}$ in the 2--10~keV
band.  If we assume this non-thermal component exists, the fit constrains
the photon index to be $\Gamma<3.0$.  With the SXI data in hand, we are
unable to say conclusively that the non-thermal component is required, only
that it is consistent with the observed spectrum.

\section{Discussion}
\label{sect:disc}

We have revealed a significant redshift of the emission lines of He-like
Fe, constraining the line-of-sight velocity to be $\sim$\,1100~km~s$^{-1}$,
or $\sim$\,800~km~s$^{-1}$ faster than the local LMC ISM.  The emission of
S\,\textsc{xv} He$\alpha$, on the other hand, shows a velocity consistent
with the radial velocity of the LMC ISM, albeit with large uncertainty,
especially when broadening is included in the model.  These results suggest
different origins of the Fe and S emission: the former is dominated by the
fast-moving ejecta and the latter by the swept-up ISM.  This interpretation
is consistent with the previous work by XMM-Newton, which revealed that the
Fe emission has a centrally-filled morphology and the S emission is found
along the outer shell \citep{Behar01}.

This interpretation hinges on our assumed underlying emission model.
Previous results from XMM-Newton \citep{Behar01} and the detection of an
Fe\,\textsc{xxvi} Ly$\alpha$ line in the Suzaku spectrum \citep{Bamba17}
suggest minor contamination from lower-energy, lower-ionization states of
Fe.  It is possible that the H-like Fe emission arises from a much hotter
plasma that does not produce He-like emission, and the Fe K complex in
question is produced by lower-temperature plasma unresolved by both the
Suzaku and Hitomi PSF.  Although L-shell lines of lower-ionization Fe were
not detected by \citep{Behar01}, it is further possible that the L-shell
energy band is dominated by the low temperature swept-up ISM component,
hindering detection of faint ejecta lines.  We are unable to conclusively
demonstrate the validity of our assumptions with existing X-ray data, and
we stress that the discussion that follows assumes the Fe K emission is
dominated by He-like Fe.

The best-fit broadening widths for both Fe K and S K, $\sigma \sim 500$
km~s$^{-1}$, greatly exceed thermal broadening at these temperatures.  It
is unclear whether the constraints on broadening are physical or somehow
related to the combination of low statistics and complicated line structure
in the thermal model.  The addition of broadening simplifies the posterior
velocity distribution without greatly changing the 90\% credible interval,
and we can speculate that if this line broadening is physical, there could
be Fe K-emitting material at a range of velocities due to bulk motion,
including very high ones.  Much better statistics at similar spectral
resolution are required to further understand the velocity structure in
both Fe K and S K.

These Fe-rich ejecta display very different line-of-sight
velocity structure compared to the O-rich ejecta explored in detail in the
optical.  The O-rich ejecta traced by [O \textsc{iii}] $\lambda 5007$
emission have an average blueshifted velocity of
$\sim\,-500$ km~s$^{-1}$ with respect to the local LMC when an elliptical
shell model is fit in projected space and velocity \citep{Morse95}.
\citet{VogtDopita2011} confirm this systematic offset, but point out that
the complicated spatial structure of the ejecta heavily biases the average
velocity of the emission as different clumps interact with the reverse
shock at different times.  The ring structure of the O-rich ejecta first
suggested by \citet{Lasker1980} and confirmed in several successive studies
is possibly accompanied by a polar jet associated with a ``run-away'' knot
and the enhanced X-ray emission along the southwestern shell
\citep{VogtDopita2011}.  It is tempting to speculate that the Fe emission
is associated with such a jet, but a more significant detection at higher
spatial resolution is required.

The lack of blue-shifted emission indicates a highly asymmetric
distribution of the Fe-rich ejecta.  Such asymmetry is seen morphologically
in the ejecta of other core-collapse SNRs, such as Cas~A
\citep{Grefenstette2017}, G292.2$+$1.8 \citep{Bhalerao2015}, and W49B
\citep{Lopez2013a}, and in the more evolved SNRs dominated by shocked
ISM/CSM but with Fe knots such as Puppis~A
\citep{Hwang2008,Katsuda2008,Katsuda2013}.  Notably, the Fe ejecta in these
remnants are not always centrally concentrated, as would
be expected in a typical core-collapse explosion.  In Cas~A, the mismatch
between the shocked Fe ejecta and more concentrated, redshifted $^{44}$Ti
has been interpreted in light of the SN explosion mechanism involving
instabilities such as SASI \citep{Grefenstette2017}.  N132D is more evolved
than Cas~A and perhaps better compared to W49B, with which it is comparable
in age.  The X-ray morphology of N132D is more symmetric than W49B, and
relatively symmetric among core-collapse SNRs in general \citep{Lopez11},
despite the obvious differences between the bright southern shell and the
blown-out northeastern region.  This symmetry could indicate a projection
effect and an axis of symmetry along the line-of-sight.  If N132D were
observed perpendicular to the direction it is, it might appear more highly
asymmetric, like W49B.  

The origin of the over-ionized plasma is not completely clear.
Interestingly, both N132D and W49B show evidence for overionization of the
Fe ejecta \citep{Ozawa2009,Bamba17}, suggesting a possible connection
between asymmetric ejecta distribution and overionization.  In addition,
recombining plasma is observed in several mixed-morphology SNRs that are
interacting with molecular clouds (e.g., \cite{Yamaguchi2009,Uchida2015});
although the mechanism responsible for the peculiar plasma conditions in
these remnants is still unclear, a possible connection is the inhomogeneous
medium into which the SNR is expanding.  In N132D, the entire southern half
of the remnant is surrounded in projection by molecular gas, with Mopra
22-m telescope CO data showing that the outer shell is sweeping through the
cloud \citep{Banas1997,Sano2015}.  This molecular gas distribution combined
with the X-ray emission morphology showing a brighter shell impinging on
the cloud in the south suggest that the shock is slowing here due to the
cloud, while the fainter shell blowing out toward the north and northeast
suggests that the shock is expanding faster here.  The detection of both
GeV emission \citep{Ackermann16} and neutral Fe K \citep{Bamba17} from
N132D further suggest that accelerated protons are interacting with the
nearby molecular cloud \citep{Bamba17}.   It is likely that N132D is
expanding into a highly inhomogeneous medium.

In W49B, the recombining plasma is detected on the west side of the remnant
whereas the molecular cloud is to the east, suggesting that the dominant
cooling mechanism producing the over-ionized plasma is rapid expansion of
the inner ejecta
\citep{Miceli2010,Lopez2013a}.  A similar density gradient is apparent in
N132D, however due to the insufficient spatial resolution of either Suzaku
or Hitomi we are unable to identify exactly where the recombining plasma is
located.  Comparing the Fe K map from XMM-Newton (figure 4a of
\cite{Behar01}) with the molecular gas map (figure 1b of \cite{Sano2015}),
we see that the Fe K peak is not in the center of the remnant nor toward
the blown-out low-density northeast region, but offset closer to the bright
southeastern shell.  Since the recombining Fe\,\textsc{xxv} He$\alpha$ is
the brightest feature seen in this spectral region, this hints that the
over-ionized plasma is located near the molecular cloud.  With the data
currently in hand, and with likely projection effects along the
line-of-sight, a firm conclusion is not possible.

It was unfortunate that the first microcalorimeter observation of a
thermally dominated SNR was not fully performed due to an attitude control
problem.  However, this short-exposure observation of N132D demonstrates
the power of high-spectral-resolution detectors by detecting clear emission
features with extremely low photon counts---a similarly short CCD
observation would not have detected these features, let alone placed
interesting constraints on the velocity.  The very low SXS background of
$\sim$\,1 event per spectral resolution element per 100 ks is also vital
for this result, and it opens the possibility of using slew observations
for similar science with similar future instruments.  For N132D, revealing
the Fe ejecta line-of-sight velocity structure, along with its detailed
spatial distribution and proper motion, is a vital step to determine its
three-dimensional velocity.  Future observations with the X-ray Astronomy
Recovery Mission (XARM) microcalorimeter, identical in performance to that
on Hitomi, will be sufficient to spatially resolve the remnant into two
regions and explore in detail the line-of-sight velocity and ionization
state for each element.  Observations with Athena \citep{AthenaWP2013} will
also be crucial to more accurately constrain the kinematics and ionization
state of this SNR.

\section{Conclusions}
\label{sect:conc}

In this paper, we have presented observations of the LMC SNR N132D taken
with Hitomi.  Using only a short, 3.7 ks observation with the SXS, we
detect emission lines of Fe\,\textsc{xxv} and S\,\textsc{xv} He$\alpha$
with only 17 and 16 counts, respectively.  Assuming a plausible emission
model and prior on the velocity broadening, the Fe line shows a redshift of
800 km~s$^{-1}$ (50--1500 km~s$^{-1}$ 90\% credible
interval) compared to the local LMC ISM, indicating that it likely arises
from highly asymmetric ejecta.  The S line is consistent with the local LMC
standard of rest, shifted by $-$65~km~s$^{-1}$ ($-$450 to
$+$435~km~s$^{-1}$ 90\% credible interval) assuming no broadening, and
likely arises from the swept-up ISM.  Longer SXI observations produce
results consistent with a recent combined Suzaku+NuSTAR spectral analysis,
including a recombining thermal plasma component responsible for the
Fe\,\textsc{xxv} He$\alpha$ emission and constraints on a non-thermal
component that dominates at high energies \citep{Bamba17}.  In addition to
this first result on SNRs with a microcalorimeter, the observations
highlight the power of high-spectral-resolution X-ray imaging instruments
in even short exposures.

\begin{trueauthors}
E.~Miller and H.~Yamaguchi led this study and wrote the final manuscript
along with S.~Katsuda, K.~Nobukawa, M.~Nobukawa, S.~Safi-Harb, and
M.~Sawada.  E.~Miller, T.~Sato, M.~Sawada, and H.~Yamaguchi performed the
SXS data reduction and analysis.  K.~Nobukawa and M.~Nobukawa performed the
SXI data reduction and analysis.  C.~Kilbourne contributed estimates and
discussion of the SXS gain uncertainty.  A.~Bamba contributed the detailed
spectral model used for both the SXS and SXI analysis.  M.~Sawada
contributed to optimizing the SXS data screening.  K.~Mori contributed
analysis of the SXI light leak.  L.~Gallo, J.~Hughes, R.~Mushotzky,
C.~Reynolds, T.~Sato, M.~Tsujimoto, and B.~Williams contributed valuable
comments on the manuscript.  The science goals of Hitomi were discussed and
developed over more than 10 years by the ASTRO-H Science Working Group
(SWG), all members of which are authors of this manuscript.  All the
instruments were prepared by joint efforts of the team.  The manuscript was
subject to an internal collaboration-wide review process.  All authors
reviewed and approved the final version of the manuscript.
\end{trueauthors}

\begin{ack}
We thank the support from the JSPS Core-to-Core Program.
We acknowledge all the JAXA members who have contributed to the ASTRO-H (Hitomi)
project.
All U.S. members gratefully acknowledge support through the NASA Science Mission
Directorate. Stanford and SLAC members acknowledge support via DoE contract to SLAC
National Accelerator Laboratory DE-AC3-76SF00515. Part of this work was performed under
the auspices of the U.S. DoE by LLNL under Contract DE-AC52-07NA27344.
Support from the European Space Agency is gratefully acknowledged.
French members acknowledge support from CNES, the Centre National d'\'{E}tudes Spatiales.
SRON is supported by NWO, the Netherlands Organization for Scientific Research.  Swiss
team acknowledges support of the Swiss Secretariat for Education, Research and
Innovation (SERI).
The Canadian Space Agency is acknowledged for the support of Canadian members.  
We acknowledge support from JSPS/MEXT KAKENHI grant numbers 15H00773, 15H00785,
15H02090, 15H03639, 15H05438, 15K05107, 15K17610, 15K17657, 16H00949, 16H06342,
16K05295, 16K05300, 16K13787, 16K17672, 16K17673, 21659292, 23340055, 23340071,
23540280, 24105007, 24540232, 25105516, 25109004, 25247028, 25287042, 25400236,
25800119, 26109506, 26220703, 26400228, 26610047, 26800102, JP15H02070, JP15H03641,
JP15H03642, JP15H03642, JP15H06896, JP16H03983, JP16K05296, JP16K05309, JP16K17667, and
JP16K05296.
The following NASA grants are acknowledged: NNX15AC76G, NNX15AE16G, NNX15AK71G,
NNX15AU54G, NNX15AW94G, and NNG15PP48P to Eureka Scientific.
H. Akamatsu acknowledges support of NWO via Veni grant.  
C. Done acknowledges STFC funding under grant ST/L00075X/1.  
A. Fabian and C. Pinto acknowledge ERC Advanced Grant 340442.
P. Gandhi acknowledges JAXA International Top Young Fellowship and UK Science and
Technology Funding Council (STFC) grant ST/J003697/2. 
Y. Ichinohe, K. Nobukawa, T. Sato, and H. Seta are supported by the Research Fellow of 
JSPS for Young Scientists.
N. Kawai is supported by the Grant-in-Aid for Scientific Research on Innovative Areas
``New Developments in Astrophysics Through Multi-Messenger Observations of Gravitational
Wave Sources''.
S. Kitamoto is partially supported by the MEXT Supported Program for the Strategic
Research Foundation at Private Universities, 2014-2018.
B. McNamara and S. Safi-Harb acknowledge support from NSERC.
T. Dotani, T. Takahashi, T. Tamagawa, M. Tsujimoto and Y. Uchiyama acknowledge support
from the Grant-in-Aid for Scientific Research on Innovative Areas ``Nuclear Matter in
Neutron Stars Investigated by Experiments and Astronomical Observations''.
N. Werner is supported by the Lend\"ulet LP2016-11 grant from the Hungarian Academy of
Sciences.
D. Wilkins is supported by NASA through Einstein Fellowship grant number PF6-170160,
awarded by the Chandra X-ray Center, operated by the Smithsonian Astrophysical
Observatory for NASA under contract NAS8-03060.

We thank contributions by many companies, including in particular, NEC, Mitsubishi Heavy
Industries, Sumitomo Heavy Industries, and Japan Aviation Electronics Industry. Finally,
we acknowledge strong support from the following engineers.  JAXA/ISAS: Chris Baluta,
Nobutaka Bando, Atsushi Harayama, Kazuyuki Hirose, Kosei Ishimura, Naoko Iwata, Taro
Kawano, Shigeo Kawasaki, Kenji Minesugi, Chikara Natsukari, Hiroyuki Ogawa, Mina Ogawa,
Masayuki Ohta, Tsuyoshi Okazaki, Shin-ichiro Sakai, Yasuko Shibano, Maki Shida, Takanobu
Shimada, Atsushi Wada, Takahiro Yamada; JAXA/TKSC: Atsushi Okamoto, Yoichi Sato, Keisuke
Shinozaki, Hiroyuki Sugita; Chubu U: Yoshiharu Namba; Ehime U: Keiji Ogi; Kochi U of
Technology: Tatsuro Kosaka; Miyazaki U: Yusuke Nishioka; Nagoya U: Housei Nagano;
NASA/GSFC: Thomas Bialas, Kevin Boyce, Edgar Canavan, Michael DiPirro, Mark Kimball,
Candace Masters, Daniel Mcguinness, Joseph Miko, Theodore Muench, James Pontius, Peter
Shirron, Cynthia Simmons, Gary Sneiderman, Tomomi Watanabe; ADNET Systems: Michael
Witthoeft, Kristin Rutkowski, Robert S. Hill, Joseph Eggen; Wyle Information Systems:
Andrew Sargent, Michael Dutka; Noqsi Aerospace Ltd: John Doty; Stanford U/KIPAC: Makoto
Asai, Kirk Gilmore; ESA (Netherlands): Chris Jewell; SRON: Daniel Haas, Martin Frericks,
Philippe Laubert, Paul Lowes; U of Geneva: Philipp Azzarello; CSA: Alex Koujelev, Franco
Moroso.

We finally acknowledge helpful comments from Mikio Morii on the statistical
analysis, and valuable comments from the anonymous referee that greatly
improved the manuscript.

\end{ack}

\end{document}